\documentclass[reqno,12pt]{article}
\pdfoutput=1
\usepackage{ae}
\usepackage[T1]{fontenc}
\usepackage[ansinew]{inputenc}
\usepackage{mathrsfs}
\usepackage{amsmath}
\usepackage{dsfont}
\usepackage{amssymb}
\usepackage{upgreek}
\usepackage{graphicx}
\usepackage{color}
\definecolor{darkblue}{cmyk}{0.9,0.9,0,0}
\definecolor{darkgreen}{rgb}{0,0.55,0}
\usepackage[colorlinks=true,linkcolor=darkblue,citecolor=darkblue,urlcolor=darkblue]{hyperref}
\usepackage{epsfig}
\usepackage{cite}
\usepackage{mathtools,slashed}
\usepackage{caption}
\usepackage{subcaption}
\usepackage{cancel}

\usepackage{amsmath,amsfonts,amssymb,amsbsy}
\usepackage{graphicx,xcolor}
\usepackage{graphics}
\usepackage{epsfig}
\usepackage{verbatim}
\usepackage{stmaryrd}
\usepackage{epsf}
\usepackage{transparent}

\usepackage{amsmath, amssymb,graphicx}
\usepackage{epsfig}
\usepackage{verbatim}
\usepackage{amsfonts}
\usepackage{ulem}
\usepackage{wrapfig}
\usepackage{tikz}
\numberwithin{equation}{section}

\usepackage{tikz}
\usetikzlibrary{arrows,shapes}
\usetikzlibrary{patterns,fadings}
\usetikzlibrary{decorations.pathmorphing}
\usetikzlibrary{decorations.markings}

\newcommand{\btp}{\begin{tikzpicture}[baseline=0pt,scale=0.9,line width=0.7pt]}
\newcommand{\btpp}{\begin{tikzpicture}[baseline=-5pt,scale=0.25,line width=0.7pt]}
\newcommand{\etp}{\end{tikzpicture}}

\def\bc{\begin{center}}
\def\ec{\end{center}}

\usepackage{epsf}

\newcommand{\be}{\begin{equation}}
\newcommand{\ee}{\end{equation}}
\newcommand{\ba}{\begin{eqnarray}}
\newcommand{\ea}{\end{eqnarray}}

\newcommand{\beaa}{\begin{eqnarray}}
\newcommand{\eeaa}{\end{eqnarray}}

\newcommand{\dbra}{\langle\!\langle}
\newcommand{\dket}{\rangle\!\rangle}
\newcommand{\T}{\mathbb{T}}
\newcommand{\D}{\mathbb{D}}

\definecolor{darkgreen}{rgb}{0.1,0.7,0.1}

\DeclareFontFamily{OT1}{pzc}{}
\DeclareFontShape{OT1}{pzc}{m}{it}{<-> s * [1.10] pzcmi7t}{}
\DeclareMathAlphabet{\mathpzc}{OT1}{pzc}{m}{it}

\def\({\left(}
\def\){\right)}
\def\[{\left[}
\def\]{\right]}

\def\<{\langle}
\def\>{\rangle}

\def\s*{\ *_{\!\!\!\!\!\!\!\!\!\,_{\,_\text{\scriptsize{sym}}}}}
\def\hs*{\ \hat{*}_{\!\!\!\!\!\!\!\!\!\,_{\,_\text{\scriptsize{sym}}}}}

\begin{document}

\thispagestyle{empty}

\renewcommand{\thefootnote}{\fnsymbol{footnote}}
\setcounter{footnote}{0}
\setcounter{figure}{0}

\begin{center}
$$$$

\vspace{1.0cm}

{\Large\textbf{\mathversion{bold}
Heavy-Heavy-Light-Light correlators \\
from holography on the worldsheet
}\par}

\vspace{1.0cm}

\vspace{1.0cm}

\textrm{Mat\'ias Colaneri.\footnote{matias.colaneri@fisica.unlp.edu.ar}, Diego H. Correa\footnote{correa@fisica.unlp.edu.ar},  Victor I. Giraldo-Rivera\footnote{vigirald@gmail.com} and \\ 
Nicol\'as A. Ivanovich \footnote{nicolas.ivanovich@fisica.unlp.edu.ar}}
\\ \vspace{1.2cm}
\footnotesize{

\textit{Instituto de F\'{\i}sica La Plata, Universidad Nacional de La Plata, \\ C.C. 67, 1900 La Plata, Argentina} \\
\texttt{} \\
}

\par\vspace{1.2cm}

\textbf{Abstract}\vspace{2mm}
\end{center}

\noindent

We study correlation functions of light local operators and  defect changing operators on the conformal line set by the $1/2$ BPS Wilson line of $\mathcal N=4$ super Yang-Mills. We consider the defect changing operators associated with geometrical and internal cusp angles, which at strong coupling have a large scaling dimension and back-react on the dual $AdS_2$ world-sheet. We set up a holographic computation of correlation functions, in which the light operators are described by fluctuations around the cusped classical string solution, so that correlators of two defect changing operators and an arbitrary number of light protected operators are obtained from the on-shell action of these fluctuations. We implement this prescription in detail for the case of certain 4-point functions, with two light and two defect changing operators.
Finally, since for small cusp angles the defect changing operators become light, we manage to derive relations satisfied by integrated correlators of light operators, valid at any coupling. We verify these relations at strong coupling for both  internal and geometrical cusps.

\vspace*{\fill}

\setcounter{page}{1}
\renewcommand{\thefootnote}{\arabic{footnote}}
\setcounter{footnote}{0}

\newpage

\section{Introduction}
\label{introduction}

The dual holographic description of a conformal line defect is that of an $AdS_2$ defect within an $AdS_{d+1}$ spacetime. Therefore, by studying the transverse fluctuations on the $AdS_2$ defect, it is possible to compute the strong coupling expression for the correlation function of excitations inserted in the conformal line, as done in \cite{Giombi:2017cqn} for the 1/2 BPS Wilson line in ${\cal N}=4$ super Yang-Mills. In  \cite{Giombi:2017cqn}, the analysis is restricted to ``light'' fields on the $AdS_2$ string world-sheet. The 8 transverse fluctuations to the string world-sheet are split into 5 massless scalar fields and 3 scalar fields with $m^2=2$ in $AdS_2$, corresponding to scalar insertions on the conformal line with scaling dimensions $\Delta=1$ and $\Delta=2$ respectively. All these insertions belong to the so-called ``displacement multiplet'' and their scaling dimensions are protected. Correlation functions of these protected operators have been studied thoroughly with a variety of theoretical tools, ranging from integrability to conformal bootstrap \cite{Drukker:2006xg,Liendo:2018ukf,Ferrero:2021bsb,Cavaglia:2021bnz,Ferrero:2023znz,Ferrero:2023gnu,Cavaglia:2022qpg,Cavaglia:2022yvv}.

Much less is known when other non-protected operators are included in the correlators.  One case of interest is that of ``heavy'' excitations associated with {\it Defect Changing Operators}, which introduce a sudden change in the direction of the Wilson line (a geometrical cusp angle $\phi$), as well as a sudden change in its internal space orientation (an internal cusp angle $\theta$). In this paper, we will elaborate on the holographic description of the insertion of two heavy defect changing operators and two or more protected light local operators. 

Interpreting a cusp as a primary operator is possible because it occurs at a point along a contour and carries a scaling dimension: the cusp anomalous dimension $\Gamma_{\rm cusp}(\phi,\theta)$. In ${\cal N}=4$ super Yang-Mills the cusp anomalous dimension can be  described  by means of integrability \cite{Drukker:2012de,Correa:2012hh} and the sum of ladder diagrams \cite{Correa:2012nk, Gromov:2015dfa} in a certain parametric limit. When 3 of such cusps are considered along a contour, their structure constants can also be determined \cite{Kim:2017sju,Cavaglia:2018lxi}. More recently, cusped line defects have also been analysed in generic conformal field theories: the conformal properties of the operators sitting at the cusp ---which are precisely the defect changing operators of interest here--- together with the cusp operator expansion, by means of which defects of general shape can be expanded in a basis of cusped defects, were established in \cite{Bianchi:2026dhj}, while non-perturbative bounds on cusp anomalous dimensions, following from unitarity, reflection positivity and the consistency of different defect geometries, were derived in \cite{Lanzetta:2026ogn,Lanzetta:2026pvr}.

What makes the defect changing operators qualitatively different from the light insertions is that their scaling dimension $\Gamma_{\rm cusp}(\phi,\theta)$ grows as $\sqrt\lambda$ at strong coupling for generic values of the cusp angles.
Having a scaling dimension of the same order as the central charge of the defect CFT$_1$, these insertions back-react on the dual world-sheet geometry, which ceases to be $AdS_2$. The classical string dual to a cusped Wilson line is known in terms of Jacobi elliptic functions \cite{Drukker:2007qr}, as is the spectrum of its quadratic fluctuations \cite{Drukker:2011za}. The insertion of light operators can still be described by fluctuations, but now around the back-reacted geometry. The main proposal of this paper is that the normalized correlator of two defect changing operators and $n$ light operators is computed holographically by the $n^{\rm th}$ functional derivative of the on-shell action of these fluctuations with respect to their boundary values.

A useful feature of this setup is that the {\it heavy} nature of the defect changing operators is not fixed, but depends continuously on the cusp angles. Since $\Gamma_{\rm cusp}(\phi,\theta)$ vanishes as the angles go to zero, the very same operators that are heavy for generic angles become light in that limit. 
This interpolation is more than a curiosity: by expanding our results in powers of the cusp angles, one obtains relations that integrated correlators of light operators must satisfy. These relations hold at any value of the coupling, even though we shall only verify them at strong coupling.

Besides the small cusp angle regimes considered in the present article, this setup admits other interesting parametric limits. For $\phi\to\pi$, the two branches of the Wilson line become anti-parallel and the configuration is governed by the quark--antiquark potential, while the limit $i\theta\to \infty$ with $\lambda e^{i\theta}$ held fixed defines the so-called ladder limit.  Finally, for $\phi=\pm\theta$, the configuration becomes supersymmetric and $\Gamma_{\rm cusp}$ vanishes identically \cite{Drukker:1999zq}. Our holographic description holds for generic $(\phi,\theta)$ and therefore provides a natural starting point to explore how heavy-light correlators behave in each of these other parametric regimes.

The paper is organized as follows. In section \ref{DCO} we review the defect changing operators associated with internal and geometrical cusp angles, their relation to the Bremsstrahlung function, and we set up the kinematics of the heavy-heavy-light-light correlators we shall study. In section \ref{HHLLLL} we present our holographic prescription and apply it to the case of two defect changing operators and two light insertions, obtaining the corresponding functions of the cross-ratio both for the purely geometrical cusp, where they can be written in closed form, and for generic cusp angles, where we provide two complementary series representations. 
In section \ref{smallgeometriccusp} we derive relations for integrated correlators of light operators that follow from the small cusp angle limit, and we verify them in the strong coupling regime. We close with a discussion of possible extensions. Three appendices collect, respectively, the structure of four-point functions of light operators and their strong coupling expressions, the details of the classical
string solution and its quadratic fluctuations and the construction of the world-sheet Green's functions.

\section{Defect Changing Operators}
\label{DCO}

Among the possible defects in a conformal field theory, a special class is that of defects of co-dimension 1, as they can act as interfaces between separate space-time regions. It is therefore possible to have different CFTs on each side of the defect. In the case of conformal line defects, such co-dimension 1 defects are point-like: they are operators pinned at a point on the line, separating two different CFT$_1$. We will refer to them as  {\it Defect Changing Operators} (DCO).

We will focus our attention on the case in which the conformal line is the Maldacena-Wilson loop \cite{Maldacena:1998im}
in ${\cal N}=4$ super Yang-Mills:
\begin{equation}
    W = \frac{1}{N}{\rm tr} P\left[ 
    \exp\left(\oint dt(i A_\mu \dot x^\mu +|\dot x|\Phi_a n^a) \right)\right]\,.
\end{equation}
When the contour is parametrized as $x^\mu(t)=\delta^\mu_0 t$ and the internal space orientation is a constant unit vector $n^a = \delta^a_6$, this is a conformal line. It is possible to excite the line by inserting local operators  and compute their correlations as follows:
\begin{equation}
    \langle\!\langle {\cal O}_1(t_1)\cdots {\cal O}_n(t_n) \rangle\!\rangle :=
    \frac{1}{N} \Big\langle {\rm tr} P\left[ {\cal O}_1(t_1)\cdots {\cal O}_n(t_n)
    \exp\left(\oint dt(i A_\mu \dot x^\mu +|\dot x|\Phi_a n^a) \right)\right]
    \Big\rangle\,.
\end{equation}
Some instances of such local operators, which will also play an important role in our analysis, are the {\it tilt} and {\it displacement} operators:
\begin{equation}
\mathbb{T}_a =  \Phi_a\,,
\qquad
\mathbb{D}_i =  i F_{it} + D_i\Phi_6\,,
\qquad a=1,\cdots, 5 \quad i=1,\cdots, 3
\end{equation}
whose scaling dimensions are protected, so that, their 2-point functions are exactly known
\begin{equation}
\langle\!\langle  \mathbb{T}_a (t_1)  \mathbb{T}_b (t_2) \rangle\!\rangle=  \frac{\gamma(\lambda) \delta_{ab}}{|t_1-t_2|^2}\,,
\qquad
\langle\!\langle \mathbb{D}_i (t_1) \mathbb{D}_j (t_2) \rangle\!\rangle =  \frac{\tilde\gamma(\lambda)\delta_{ij}}{|t_1-t_2|^4}\,.
\label{2ptfunctions}
\end{equation}
The 't Hooft coupling dependent normalizations $\gamma(\lambda)$ and $\tilde\gamma(\lambda)$ are tied to the Bremsstrahlung function $B(\lambda)$ \cite{Correa:2012at}, as we will review in what follows. Conformal symmetry also constrains the correlation functions containing more than two of these light protected operators. For example, 4-point functions are determined up to functions of the unique independent cross-ratio. Furthermore, these functions must satisfy certain Bose symmetry relations (see Appendix \ref{app:4pt}).

In addition to these local operators, one can consider the aforementioned DCOs, by introducing abrupt changes in the direction or the internal space orientation of the Maldacena-Wilson line. For example, we will use ${\cal F}_a(t,\theta)$ to denote an internal cusp angle $\theta$ in the $a^{\rm th}$ internal space direction,
\begin{eqnarray}
\langle\!\langle 
{\cal F}_a (t_1,\theta) 
{\cal F}_a (t_2,-\theta) 
\rangle\!\rangle
&\! = \! &
 \frac{1}{N}\Big\langle {\rm tr} P\left[ 
    \exp\left(\int_{-\infty}^{t_1}dt(i A_\mu \dot x^\mu +|\dot x|\Phi_a n^a) \right)
    \right. \times \nonumber
    \\
    & &  \hspace{-2.6cm} \left.
     \exp\left(\int_{t_1}^{t_2}dt(i A_\mu \dot x^\mu +|\dot x|\Phi_a {n'}^a) \right)
     \exp\left(\int_{t_2}^{\infty}dt(i A_\mu \dot x^\mu +|\dot x|\Phi_a n^a) \right)
    \right]\Big\rangle
    \label{2Fs}
 \end{eqnarray}
with $\delta_{ab} n^a {n'}^b=\cos\theta$. 
The expectation value \eqref{2Fs} exhibits UV divergences  localized at the two cusp points. Once these divergences are regularized, it takes the form
\begin{equation}
    \langle\!\langle 
{\cal F}_a (t_1,\theta) 
{\cal F}_a (t_2,-\theta) 
\rangle\!\rangle \propto \left(\frac{\epsilon}{|t_{12}|}\right)^{2\Gamma_{\rm cusp}(\theta)}\,,
\label{FFcusp}
\end{equation}
where $t_{ij}=t_i-t_j$ and $\epsilon$ is an ultraviolet cutoff. This allows us to interpret the cusps as operators located at $t_1$ and $t_2$, whose scaling dimension is
the usual internal cusp anomalous dimension $\Gamma_{\rm cusp}(\theta)$ \cite{Kim:2017sju}. We stress that \eqref{FFcusp} is nothing but the two-point function of a primary operator of dimension $\Gamma_{\rm cusp}(\theta)$. 
 
In the strong coupling limit, and for generic values of $\theta$, $\Gamma_{\rm cusp}$ scales as $\sqrt\lambda$, the same order as the central charge of the dCFT$_1$. Thus, the cusp operator constitutes an example of a heavy operator that will back-react on the dual $AdS_2$ geometry. 

In the small cusp angle limit, we can represent these DCOs with integrals of light operators. Consider, for instance, the case $n^a = \delta^a_6$ and ${n'}^a = \cos\theta\delta^a_6+\sin\theta\delta^a_5$. Then,
\begin{equation}
    \langle\!\langle 
{\cal F}_5 (0,\theta) 
{\cal F}_5 (L,-\theta) 
\rangle\!\rangle
\label{FFasexpvev0}
=
\big\langle\!\big\langle 
\exp\left(\int_0^L \!\!dt\,(\Phi_a(t){n'}^a-\Phi_a(t){n^a})\right)
\big\rangle\!\big\rangle\,.
\end{equation}
For a small cusp angle
\begin{equation} 
\Phi_a(t){n'}^a-\Phi_a(t){n^a} = \theta \Phi_5(t)-\frac{\theta^2}2 \Phi_6(t)
+{\cal O}{(\theta^3)}\,,
\end{equation}
and the expansion of \eqref{FFasexpvev0} in powers of $\theta$ gives a series of integrals of correlators of $\Phi_5$ and $\Phi_6$. For example,
\begin{eqnarray}
\langle\!\langle 
{\cal F}_5 (0,\theta) 
{\cal F}_5 (L,-\theta)
\rangle\!\rangle
&\! = \! & 1 + \theta \int_{0}^{L}\!\!dt\, \langle\!\langle 
\Phi_5 (t) \rangle\!\rangle 
- \frac{\theta^2}{2} \!\!\int_{0}^{L}\!\!dt \, 
    \langle\!\langle 
\Phi_6 (t) \rangle\!\rangle 
\nonumber
\\
&& 
   + \frac{\theta^2}{2} \!\!\int_{0}^{L}\!\!dt\int_{0}^{L}\!dt'\, 
    \langle\!\langle 
\Phi_5 (t) \Phi_5 (t') \rangle\!\rangle 
+{\cal O}(\theta^3)\,.
    \label{2Fssapp0}
 \end{eqnarray}

To evaluate such integrated correlators, one needs to know how the scalar fields $\Phi_5$ and $\Phi_6$ are related to primary operators of the CFT$_1$. As already mentioned, the scalar field $\Phi_5$ is in correspondence with one of the components of the tilt operator $\mathbb{T}_a$, a protected primary operator that  transforms in the $\mathbf{5}$ of $SO(5)$. Thus,
\begin{equation}
\langle\!\langle 
\Phi_5(t) \rangle\!\rangle = \langle\!\langle 
\mathbb{T}_5(t) \rangle\!\rangle = 0\,, \qquad   
\langle\!\langle 
\Phi_5(t) \Phi_5(t') \rangle\!\rangle = \langle\!\langle 
\mathbb{T}_5(t) \mathbb{T}_5(t') \rangle\!\rangle = \frac{\gamma}{(t'-t)^2}\,,   
\end{equation}
where $\gamma$ is a normalization.
On the other hand, the longitudinal scalar field $\Phi_6$ is a singlet of $SO(5)$ and it corresponds, up to a UV scaling factor, to the composite two-particle operator \cite{Giombi:2017cqn}
\begin{equation}
\Phi_6(t) = \zeta \epsilon^{\Delta-1}\sum_{a=1}^5 \mathbb{T}_a(t-\tfrac{\epsilon}{2})\mathbb{T}_a(t+\tfrac{\epsilon}{2})\,, 
\label{Phi6proposal}
\end{equation}
where $\Delta$ is the scaling dimension of the unprotected two-particle operator. 
In what follows we work to leading order in the strong coupling limit, and then 
$\Delta = 2 + {\cal O}(\frac{1}{\sqrt\lambda})$. The constant of proportionality $\zeta$ can be adjusted to cancel a power-law divergence that comes from the double integral in the second line of \eqref{2Fssapp0}.

The decomposition of the operator $\Phi_6(t)$ into primaries and descendants can be obtained from the 4-point correlator
$\langle\!\langle 
\mathbb{T}_a(t-\tfrac{\epsilon}{2})\mathbb{T}_a(t+\tfrac{\epsilon}{2}) \mathbb{T}_b(t'-\tfrac{\epsilon}{2})\mathbb{T}_b(t'+\tfrac{\epsilon}{2}) \rangle\!\rangle$. To leading order in the strong coupling limit, this is simply given by disconnected Witten diagrams, from which it is straightforward to infer \cite{Giombi:2017cqn}
\begin{equation}
  \Phi_6(t) =   5 \gamma \zeta \left(\frac{\mathds{1}}{\epsilon}
+ \epsilon\sqrt{\frac{2}{5}}{ O}_{\Delta=2}(t)+ {\cal O}(\epsilon^3)\right)\,,
\end{equation}
where ${ O}_{\Delta=2}(t)$ is unit-normalized. Thus,
\begin{equation}
\langle\!\langle 
{\cal F}_5 (0,\theta) 
{\cal F}_5 (L,-\theta)
\rangle\!\rangle
= 1 -  \theta^2\left(  \frac{5\gamma\zeta}{2} \frac{L}{\epsilon}-\gamma \frac{L}{\epsilon}+\gamma\log\left(\frac{L}{\epsilon}\right) +\gamma +{\cal O}(\epsilon)
\right)   +{\cal O}(\theta^3)\,, \label{2Fssap}
\end{equation}
where we have used that 
\begin{equation}
 \int_{0}^{L}\!dt\, \int_{0}^{L}\!dt'\, 
   \frac{1}{(t'-t)^2} 
   =  
   2\frac{L}{\epsilon} -
   2\log\left(
   \frac{L}{\epsilon}\right) - 2 +{\cal O}(\epsilon)\,.
   \label{integrated2pt}
\end{equation}
Notice we have identified the UV cutoffs in \eqref{Phi6proposal} and \eqref{integrated2pt}. With that choice the power-law divergent terms cancel each other by fixing $\zeta=\frac25$.

Using the  small angle limit of the  internal cusp anomalous dimension \cite{Correa:2012at}
\begin{equation}
    \Gamma_{\rm cusp} = \theta^2 B + 
    {\cal O}(\theta^4)\,,
\end{equation}
where $B$ is the so-called Bremsstrahlung function, we can re-express \eqref{FFcusp} as 
\begin{equation}
    \langle\!\langle 
{\cal F}_5 (0,\theta) 
{\cal F}_5 (L,-\theta) 
\rangle\!\rangle \propto 
1-2  \theta^2 B(\lambda) \log\left(
   \frac{L}{\epsilon}\right)+
{\cal O}(\theta^4)\,.
\label{FFcuspexpa}
\end{equation}
In order to match this with \eqref{2Fssap} one has to relate the normalization of the tilt operators with the Bremsstrahlung function 
\begin{equation}
\gamma = 2B(\lambda)\,.
\label{Ph5Phi6normalizations}
\end{equation}
By relating this normalization with a localization result, the Bremsstrahlung function is exactly determined \cite{Correa:2012at}. In the large $N$ limit the Bremsstrahlung function becomes
\begin{equation}
   B(\lambda) = \frac{\sqrt{\lambda}}{4\pi^2}\frac{I_2(\sqrt\lambda)}{I_1(\sqrt\lambda)} \,.
\end{equation}
In principle, one can proceed with a similar analysis to higher orders in $\theta$, as done in \cite{Cavaglia:2022yvv}, which would involve integrated correlators with more points.

In a similar fashion, we can define a DCO associated with a geometrical cusp angle $\phi$. We now use ${\cal C}_i(t,\phi)$ to denote a geometrical cusp in the $i^{\rm th}$ spatial direction,
\begin{eqnarray}
\langle\!\langle 
{\cal C}_i (t_1,\phi) 
{\cal C}_i (t_2,-\phi) 
\rangle\!\rangle
&\! = \! &
 \frac{1}{N}\Big\langle {\rm tr} P\left[ 
    \exp\left(\int_{-\infty}^{t_1}dt(i A_\mu \dot x^\mu +|\dot x|\Phi_a n^a) \right)
    \right. \times \nonumber
    \\
    & &  \hspace{-2.6cm} \left.
\exp\left(\int_{t_1}^{t_2}dt(i A_\mu \dot {x'}^\mu +|\dot {x'}|\Phi_a {n}^a) \right)
\exp\left(\int_{t_2}^{\infty}dt(i A_\mu \dot x^\mu +|\dot x|\Phi_a n^a) \right)
    \right]\Big\rangle\,.
    \label{2Cs}
 \end{eqnarray}
We can also consider  this 2-point function in the small cusp angle limit. We take $x^\mu = \delta^\mu_0 t$ and ${x'}^\mu = t(\cos\phi\delta^\mu_0+\sin\phi\delta^\mu_1) = \delta^\mu_0 t +\phi \delta^\mu_1 t -\tfrac{1}{2}\phi^2\delta^\mu_0 t+ {\cal O}(\phi^3)$. The expansion in powers of the geometrical cusp angle is a bit more laborious as one has to expand not only the velocity $\dot x^\mu$ but also the position at which the fields are evaluated, and then reorganise the results through integrations by parts. One gets in this case,
\begin{eqnarray}
\langle\!\langle 
{\cal C}_1 (0,\phi) 
{\cal C}_1 (L,-\phi) 
\rangle\!\rangle
&\!=\!&
1 +\phi \int_{0}^{L}\!\!dt\, \langle\!\langle 
\mathbb{D}_1  (t) \rangle\!\rangle t
+\frac{\phi^2}{2} \!\!\int_{0}^{L}\!\!dt \, 
    \langle\!\langle 
\Phi_6 (t) \rangle\!\rangle 
+\frac{\phi^2}{2} \!\!\int_{0}^{L}\!\!dt \, 
    \langle\!\langle 
D_1 \mathbb{D}_1  (t) \rangle\!\rangle t^2
\nonumber 
\\
&& + \frac{\phi^2}{2} \int_{0}^{L}\!dt\, \int_{0}^{L}\!dt'\, 
    \langle\!\langle 
\mathbb{D}_1 (t) \mathbb{D}_1 (t') \rangle\!\rangle t\, t' 
+{\cal O}(\phi^3)\,.
    \label{2Css}
 \end{eqnarray}
As the displacement $\mathbb{D}_i$ is a protected primary operator with scaling dimension 2,
\begin{equation}
\langle\!\langle 
\mathbb{D}_1(t) \rangle\!\rangle = 0\,, \qquad   
\langle\!\langle 
\mathbb{D}_1(t) \mathbb{D}_1(t') \rangle\!\rangle = \frac{\tilde \gamma}{(t'-t)^4}\,.   
\end{equation}
We already know how to express $\Phi_6$ in terms of primary fields of the CFT$_1$, but also need to express the covariant derivative of the displacement $D_1 \mathbb{D}_1$. Guided by its SO(3) quantum numbers, and in analogy with \eqref{Phi6proposal}, we propose the following composite two-particle operator:
\begin{equation}
D_1 \mathbb{D}_1(t) = \eta\, \epsilon^{\Delta-3} \mathbb{D}_1(t-\tfrac{\epsilon}{2})\mathbb{D}_1(t+\tfrac{\epsilon}{2})\,.
\label{D1D1proposal}
\end{equation}
In the strong coupling limit, the unprotected scaling dimension of the  two-particle operator is now $\Delta = 4 + {\cal O}(\frac{1}{\sqrt\lambda})$. As before, the constant of proportionality can be adjusted to cancel power-law divergences and the matching of logarithmic divergences fixes the normalization of the displacement operator,
\begin{equation}
\tilde\gamma = 12B(\lambda)\,.
\label{D1normalization}
\end{equation}
It is important to emphasize that the identifications \eqref{Ph5Phi6normalizations} and \eqref{D1normalization} are valid to all orders in the coupling constant expansion.

A more general DCO is defined by letting a geometrical and an internal cusp occur at the same point. We shall refer to it as ${\cal C}_{i,a}(t,\phi,\theta)$. 

\subsection{Heavy-Heavy-Light-Light correlators}

So far, we have reviewed the computation of two-point functions of DCOs. Naturally, we can also consider correlators that mix local operators $\mathbb{T}_a$ and $\mathbb{D}_i$ with DCOs ${\cal F}_a$ and ${\cal C}_i$. In the next section, we discuss the holographic prescription to compute --in the strong coupling limit-- correlators with two DCOs and various local operators.

Since the DCOs are primary operators of scaling dimension $\Gamma_{\rm cusp}$, the correlation function of two tilt operators and two internal cusp DCOs should take the following form,
 \begin{equation}
    \langle\!\langle \mathbb{T}_1(t_1) \mathbb{T}_1(t_2)
{\cal F}_5 (t_3,\theta) 
{\cal F}_5 (t_4,-\theta) 
\rangle\!\rangle = \frac{2B\, {\cal N}_{\cal F}(\theta)\, f(u,\theta)}{t_{12}^2\left({|t_{34}|}/{\epsilon}\right)^{2\Gamma_{\rm cusp}(\theta)}}\,,
\label{2T2F}
\end{equation}
for some function $f(u,\theta)$ of the unique independent cross-ratio
\begin{equation}
    u = \frac{t_{12} t_{34}}{t_{13}t_{24}}\,.
\end{equation}
In \eqref{2T2F}, the normalizations $2B$ and ${\cal N}_{\cal F}(\theta)$ of two-point correlators of tilts and DCOs have been included for later convenience.

As the CFT$_1$ is different at each side of the DCOs, we will also consider the ordering
\begin{equation}
    \langle\!\langle \mathbb{T}_1(t_1) 
{\cal F}_5 (t_2,\theta) 
\mathbb{T}_1(t_3)
{\cal F}_5 (t_4,-\theta) 
\rangle\!\rangle = \frac{2B\, {\cal N}_{\cal F}(\theta)\, \tilde f(u,\theta)}{t_{13}^2\left({|t_{24}|}/{\epsilon}\right)^{2\Gamma_{\rm cusp}(\theta)}}\,,
\label{2T2Fb}
\end{equation}
which is expected to be physically inequivalent to \eqref{2T2F}, since here the two tilts are inserted in different defect CFTs.

For later reference, we also quote the analogous formulae in the case of geometric cusp DCOs,
 \begin{align}
\langle\!\langle \mathbb{T}_1(t_1) \mathbb{T}_1(t_2)
{\cal C}_1 (t_3,\phi) 
{\cal C}_1 (t_4,-\phi) 
\rangle\!\rangle = &\, \frac{2B\, {\cal N}_{\cal C}(\phi)\, g(u,\phi)}{t_{12}^2\left({|t_{34}|}/{\epsilon}\right)^{2\Gamma_{\rm cusp}(\phi)}}\,,
\label{TTCC}
\\
 \langle\!\langle \mathbb{T}_1(t_1) {\cal C}_1 (t_2,\phi)\mathbb{T}_1(t_3) 
{\cal C}_1 (t_4,-\phi) 
\rangle\!\rangle = &\, \frac{2B\, {\cal N}_{\cal C}(\phi)\, \tilde{g}(u,\phi)}{t_{13}^2\left({|t_{24}|}/{\epsilon}\right)^{2\Gamma_{\rm cusp}(\phi)}}\,.
\label{TCTC}
\end{align}

\section{Holographic prescription for HHL$\cdots$L correlators}
\label{HHLLLL}

For definiteness, let $J_\alpha(t)$ be the sources of the light local operators discussed above, which we collectively denote by ${\cal O}_\alpha(t)$ (${\cal O}_\alpha \equiv \mathbb{D}_\alpha$ for $\alpha=1\cdots3$ and ${\cal O}_\alpha \equiv \mathbb{T}_{\alpha-3}$ for $\alpha=4\cdots8$). Analogously, $K_\alpha(t)$ are the sources of the heavy defect changing operators, which we collectively denote by  ${\cal C}_\alpha(t)$ (${\cal C}_\alpha$ is the geometric cusp DCO for $\alpha=1\cdots3$ and ${\cal C}_\alpha \equiv {\cal F}_{\alpha-3}$ for $\alpha=4\cdots8$).

The formal definition of the generating functional reads
\begin{equation}
 Z[J,K] = \left\langle\!\!\left\langle
 e^{- \int dt ( J_\alpha(t) {\cal O}_\alpha(t)+K_\alpha(t){\cal C}_\alpha(t))}
 \right\rangle\!\!\right\rangle\,,
\end{equation}
and from it, one can compute correlation functions by performing functional derivatives. For example
\begin{equation}
\left\langle\!\langle
 {\cal C}_{\alpha_1}(t_1) {\cal C}_{\alpha_2}(t_2)
 {\cal O}_{\alpha_3}(t_3)\cdots {\cal O}_{\alpha_n}(t_{n})
 \right\rangle\!\rangle
  = 
  \left.\frac{(-1)^{n}\, \delta^{n}Z[J,K]}{\delta K_{\alpha_1}(t_1)\delta K_{\alpha_2}(t_2)\delta J_{\alpha_3}(t_3)\cdots \delta J_{\alpha_n}(t_{n})} \right|_{J,K=0}
 \,,
\end{equation}

Providing a holographic prescription to compute $Z[J,K]$ appears to be very challenging. We can nevertheless give the prescription for certain specific cases. $Z[J_\alpha,0]$ is a generating functional for correlators that contain only light operators ${\cal O}_{\alpha_i}(t_i)$. In the case in which the conformal line is the 1/2 BPS Wilson line in ${\cal N}=4$ super Yang-Mills, the holographic prescription for the light operators has been given in \cite{Giombi:2017cqn}. In that case, $Z[J_\alpha,0]$ is the partition function for the 8 transverse fluctuations $\varphi_\alpha$ around the classical open string ending on a line at the boundary of $AdS_5\times S^5$, 
\begin{equation}
 Z[J_\alpha,0] =  \int {\cal D} \varphi_\alpha\,   
 e^{-\sqrt\lambda {\rm Area} - S_{\rm fluc}[J_\alpha]}\,.
 \label{ZJ0}
\end{equation}
For the straight line, the classical world-sheet is (Euclidean) $AdS_2$, whose
regularized area is vanishing. The action for the fluctuations  $S_{\rm fluc}$ is that of 8 scalar fields in $AdS_2$. The sources $J_\alpha(t)$ are the boundary values for the scalar fields imposed through Dirichlet boundary conditions. In the large $\lambda$ limit, the path integral in \eqref{ZJ0} is reduced to the on-shell evaluation of $S_{\rm fluc}$. The order to which the expansion in fluctuations has to be carried out depends on the number of light operators included in the correlator. For computing 2-point correlators, it is enough to keep $S_{\rm fluc}$ up to quadratic order in the fluctuations,
\begin{equation}
S_{\rm fluc} = \frac{\sf k}{2}\int dt \int dz \sqrt{g} \left(\partial^\mu\varphi_\alpha \partial_\mu\varphi_\alpha +m_\alpha^2 \varphi_\alpha^2+ {\cal O}(\varphi^4)\right) \,,
\label{Sflucquadratic}
\end{equation}
with $m_\alpha^2 = 2$ for $\alpha=1,2,3$ and $m_\alpha^2 = 0$ for $\alpha=4,\cdots 8$. Working with Poincar\'e coordinates
\begin{equation}
ds^2 = \frac{dt^2+dz^2}{z^2}\,,
\end{equation}
the field configurations near the boundary for $z\to 0$ are
\begin{alignat}{2} 
\varphi_\alpha(z,t) & \simeq z^{-1} A_\alpha(t) + z^{2} B_\alpha(t)\,,& \quad & \quad \text{for } \alpha=1,2,3
\\ 
\varphi_\alpha(z,t) & \simeq A_\alpha(t) + z^{1} B_\alpha(t)\,, & \quad & \quad \text{for } \alpha=4,\cdots 8\,.
\end{alignat} 
Fixing $A_\alpha(t) = J_\alpha(t)$ as the boundary condition, one obtains correlation functions for the displacement and tilt operators, whose scaling dimensions are $\Delta=2$ and $\Delta=1$ respectively.
Notice that we have included a constant ${\sf k}$ in the normalization of the scalar fields. This constant must be adjusted to reproduce the normalization of the displacement and tilt operators obtained from a localization result.

From \eqref{ZJ0}, the 2-point correlators are  given by
\begin{equation}
\left\langle\!\langle
 {\cal O}_{\alpha_1}(t_1) {\cal O}_{\alpha_2}(t_2) \right\rangle\!\rangle
  = \left.\frac{\delta^{2}Z[J,0]}{\delta J_{\alpha_1}(t_1)\delta J_{\alpha_2}(t_2)} \right|_{J=0}
  = -\left.\frac{\delta^{2}S_{\rm fluc}[J]}{\delta J_{\alpha_1}(t_1)\delta J_{\alpha_2}(t_2)} \right|_{J=0}
 \,,
\end{equation}
which gives
\begin{alignat}{2} 
\left\langle\!\langle
 {\cal O}_{\alpha_1}(t_1) {\cal O}_{\alpha_2}(t_2) \right\rangle\!\rangle & = \frac{6\, {\sf k}}{\pi}\frac{\delta_{\alpha_1\alpha_2}}{t_{12}^4}\,,& \quad & \quad \text{for } \alpha=1,2,3
\\ 
\left\langle\!\langle
 {\cal O}_{\alpha_1}(t_1) {\cal O}_{\alpha_2}(t_2) \right\rangle\!\rangle &  =
 \frac{{\sf k} }{\pi} \frac{\delta_{\alpha_1\alpha_2}}{t_{12}^2}
 \,, & \quad & \quad \text{for } \alpha=4,\cdots 8\,.
\end{alignat}
These correlators should match the normalizations of the displacement and the tilt operators in the strong coupling limit
\begin{equation}
12B(\lambda) = \frac{3\sqrt\lambda} {\pi^2} +{\cal O}(\lambda^0)\,, \qquad 2B(\lambda) = \frac{\sqrt\lambda}{2\pi^2} +{\cal O}(\lambda^0)\,.  
\end{equation}
From this, one infers that
\begin{equation}
{\sf k} = \frac{\sqrt\lambda}{2\pi} +{\cal O}(\lambda^0) \,.
\label{knorma}
\end{equation}
This is the strong coupling limit of the normalization of the scalar fields, and it can receive corrections when subleading terms of the strong coupling expansion are included \cite{Castellani:2026wwh}.

We now turn to the functional derivatives with respect to DCO sources. To begin with, note that
\begin{equation}
\left.\frac{\delta Z[J,K]}{\delta K(t_1)} \right|_{K=0} = 0\,,
\end{equation}
as correlation functions with a single DCO vanish. Thus, the first non-trivial functional derivatives with respect to DCO sources are of second order. These can be used to compute correlation functions with two DCO insertions. These DCOs are heavy operators, with a scaling dimension of order $\sqrt\lambda$ in the strong coupling limit of the 1/2 BPS Wilson line in ${\cal N}=4$ super Yang-Mills \cite{Giombi:2017cqn,Correa:2012at,Drukker:2007qr}.  Such heavy operators back-react on the world-sheet geometry, which will cease to be just $AdS_2$. 

Therefore, to compute HHL$\cdots$L correlators, we can just consider fluctuations around the new classical world-sheet geometry. For convenience, we shall consider the case in which the two heavy DCOs are located at $t=0$ and  $t=\infty$ respectively. 
\begin{eqnarray}
    \left.\frac{\delta^2 Z[J,K]}{\delta K(\infty) \delta K(0)} \right|_{K=0} =
     \int {\cal D} \varphi_\alpha\,    e^{-\sqrt\lambda {\rm Area}(\phi,\theta)- S_{\rm fluc}^{(\phi,\theta)}[J_\alpha]}\,
     \label{partitionthetaphi}
\end{eqnarray}
The new classical world-sheet geometry is that of the open string dual to a Wilson line with geometrical and internal cusp angles $\phi$ and $\theta$, which is analytically known in terms of Jacobi elliptic functions \cite{Drukker:2007qr}. In Appendix \ref{app:fluc} we review this solution as well as the spectrum of quadratic fluctuations derived in 
\cite{Drukker:2011za}.

In the strong coupling limit, the functional integral in \eqref{partitionthetaphi} is dominated by the on-shell evaluation of 
$S_{\rm fluc}^{(\phi,\theta)}[J_\alpha]$. The factor $e^{-\sqrt\lambda {\rm Area}(\phi,\theta)}$ is just $\left\langle\!\langle
 {\cal C}_{\alpha_1}(0) {\cal C}_{\alpha_2}(\infty)
 \right\rangle\!\rangle$ to leading order in the strong coupling expansion. Dividing by it
\begin{equation}
\frac{\left\langle\!\langle
 {\cal C}_{\alpha_1}(0) {\cal C}_{\alpha_2}(\infty)
 {\cal O}_{\alpha_3}(t_3)\cdots {\cal O}_{\alpha_n}(t_{n})
 \right\rangle\!\rangle}{\left\langle\!\langle
 {\cal C}_{\alpha_1}(0) {\cal C}_{\alpha_2}(\infty)
 \right\rangle\!\rangle}
  = 
  \left.\frac{(-1)^{n}\, \delta^{n-2}\left(e^{- S_{\rm fluc}^{(\phi,\theta)}[J_\alpha]}\right)}{\delta J_{\alpha_3}(t_3)\cdots \delta J_{\alpha_n}(t_{n})} \right|_{J=0}
 \,.
 \label{proposal}
\end{equation}
Eq. \eqref{proposal} is the central proposal of our paper. It can be used to compute the indicated  correlation function in the strong coupling limit and for an arbitrary number of light operators. For $n-2$ light operators, the action $S_{\rm fluc}^{(\phi,\theta)}$ has to be taken to $(n-2)^{\rm th}$ order in the fluctuations. In what follows we focus on the cases with two light operators in addition to the two DCOs, 
\begin{equation}
\frac{\left\langle\!\langle
 {\cal C}_{\alpha_1}(0) {\cal C}_{\alpha_2}(\infty)
 {\cal O}_{\alpha_3}(t_3) {\cal O}_{\alpha_4}(t_{4})
 \right\rangle\!\rangle}{\left\langle\!\langle
 {\cal C}_{\alpha_1}(0) {\cal C}_{\alpha_2}(\infty)
 \right\rangle\!\rangle}
  = 
  -\left.\frac{\delta^{2} S_{\rm fluc}^{(\phi,\theta)}[J_\alpha]}{\delta J_{\alpha_3}(t_3)\delta J_{\alpha_4}(t_{4})} \right|_{J=0}\,,
 \label{proposalHHLL}
\end{equation}

For these Heavy-Heavy-Light-Light correlators it suffices to work with an action quadratic in the fluctuations, which takes the form of 8 non-minimally coupled scalar fields on the world-sheet. Its induced metric is
\begin{equation}
ds^2_{\rm ind} = \frac{1-k^2}{{\rm cn}^2(\sigma,k^2)}\left(d\tau^2+d\sigma^2\right)\,,\qquad {\rm for \ }
-\mathbb{K}(k^2)<\sigma< \mathbb{K}(k^2)\,.
\label{induced}
\end{equation}
which is asymptotically $AdS_2$ as $\sigma\to\pm \mathbb{K}(k^2)$.  Up to quadratic order
\begin{eqnarray}
    S_{\rm fluc}^{(\phi,\theta)} &\! =\! & \frac{\sqrt\lambda}{4\pi}\int \! d\tau\! \int^{\mathbb{K}}_{-\mathbb{K}}\! d\sigma\,\sqrt{g} \left(\partial^\mu\varphi_\alpha \partial_\mu\varphi_\alpha + M_{\alpha\beta}(\sigma) \varphi_\alpha \varphi_\beta + A(\sigma)(\varphi_3\partial_{\sigma}\varphi_8-\varphi_8\partial_{\sigma}\varphi_3)\right) 
    \nonumber
    \\
    &&  + {\cal O}(\varphi^4) \,.
\label{Sflucquadraticcusp}
\end{eqnarray}
The detailed expressions for the coupling $M_{\alpha\beta}(\sigma)$ and $A(\sigma)$ are given in Appendix \ref{app:fluc}. They all depend on two constants $(k,b)$ that parametrize the two cusp angles $(\phi,\theta)$.  We use the same normalization as for the straight line case.

Among all the possible fluctuations, four of them have relatively simple equations of motion
\begin{equation}
    \nabla_\mu\nabla^\mu \varphi_\alpha + \frac{1}{\sqrt{g}}  \overbrace{\frac{1-k^2(2+b^2)}{1+b^2}}^{M^2}\varphi_\alpha = 0\,,
    \qquad\text{for \ \ } \alpha=4,5,6,7\,.
    \label{simpleeom}
\end{equation}
We emphasize that the second term in \eqref{simpleeom} is not a mass term. Due to the presence of the $\frac{1}{\sqrt{g}}$ factor, this is vanishing as one approaches the boundaries with $\sigma\to\pm \mathbb{K}(k^2)$. Thus, the expansions of these fields near the boundary are like those of massless fields in $AdS_2$,
\begin{alignat}{2} 
\varphi_\alpha(\sigma,\tau) & \simeq A^{\rm L}_\alpha(\tau) + (\sigma+\mathbb{K}(k^2)) B^{\rm L}_\alpha(\tau)\,,& \quad & \quad \text{for } \sigma\to -\mathbb{K}(k^2)\,,
\\ 
\varphi_\alpha(\sigma,\tau) & \simeq A^{\rm R}_\alpha(\tau) - (\sigma-\mathbb{K}(k^2)) B^{\rm R}_\alpha(\tau)\,,& \quad & \quad \text{for } \sigma\to +\mathbb{K}(k^2)\,.
\end{alignat}

Now, since we are working with a conformally flat metric \eqref{induced}, the equations of motion \eqref{simpleeom} become
\begin{equation}
   \frac{1}{\sqrt{g}}\left(\partial^2_{\tau} \varphi_\alpha + \partial^2_{\sigma} \varphi_\alpha + M^2 \varphi_\alpha\right) = 0 \,,
\end{equation}
which are easily solved in terms of modes of the form $e^{\pm  W\sigma}e^{i\omega\tau}$ provided $W = \sqrt{\omega^2-M^2}$.

On-shell, the variation of the action \eqref{Sflucquadraticcusp} reduces  to boundary terms
\begin{equation}
\delta S_{\rm fluc}^{(\phi,\theta)} =
-\frac{\sqrt\lambda}{2\pi} \int \!d\tau \left(\delta A^{\rm L}_{\alpha} B^{\rm L}_{\alpha}+\delta A^{\rm R}_{\alpha} B^{\rm R}_{\alpha}\right)\,,  
\end{equation}
and from this, the HHLL and HLHL cases are given by
\begin{eqnarray}
-\left.\frac{\delta^{2} S_{\rm fluc}^{(\phi,\theta)}}
    {\delta A^{\rm L}_{\alpha_3} (\tau_3)
     \delta A^{\rm L}_{\alpha_4} (\tau_4)} \right|_{A=0} &\! = &  
     \frac{\sqrt\lambda}{2\pi} \, \frac{\delta B^{\rm L}_{\alpha_4} (\tau_4)}{\delta A^{\rm L}_{\alpha_3} (\tau_3)}\,,    
     \\
     -\left.\frac{\delta^{2} S_{\rm fluc}^{(\phi,\theta)}}{\delta A^{\rm R}_{\alpha_3} (\tau_3)
     \delta A^{\rm L}_{\alpha_4} (\tau_4)} \right|_{A=0} &\! = &  
     \frac{\sqrt\lambda}{2\pi}\, \frac{\delta B^{\rm L}_{\alpha_4} (\tau_4)}{\delta A^{\rm R}_{\alpha_3} (\tau_3)}\,,    
\end{eqnarray}
We begin by writing the superposition of Fourier modes that satisfy
\begin{equation}
\varphi_\alpha(+\mathbb{K}(k^2),\tau) =   A^{\rm R}_{\alpha} (\tau)\,,
\qquad 
\varphi_\alpha(-\mathbb{K}(k^2),\tau) =   A^{\rm L}_{\alpha} (\tau)\,,
\label{boundaryvalues}
\end{equation}
which can be put as
\begin{eqnarray}
\varphi_\alpha(\sigma,\tau) &\!=\!&  
\int\limits_{-\infty}^{\infty}\!
\frac{d\omega}{\sqrt{2\pi}}
\left[\frac{{\tilde A^{\rm R}_{\alpha}}(\omega)}{{\sinh}(2W\mathbb{K})}-{\tilde A^{\rm L}_{\alpha}}(\omega)\coth(2W\mathbb{K})\right]\sinh[W(\mathbb{K}+\sigma)]e^{i\omega\tau}
\nonumber
\\
&& +  
\int\limits_{-\infty}^{\infty}\!
\frac{d\omega}{\sqrt{2\pi}}
\,{\tilde A^{\rm L}_{\alpha}}(\omega)\cosh[W(\mathbb{K}+\sigma)]e^{i\omega\tau}
\end{eqnarray}
and from this
\begin{equation}
B^{\rm L}_{\alpha} (\tau)
=
\int\limits_{-\infty}^{\infty}\!
\frac{d\omega}{\sqrt{2\pi}}
\left[\frac{{\tilde A^{\rm R}_{\alpha}}(\omega)}{{\sinh}(2W(\omega)\mathbb{K})}-{\tilde A^{\rm L}_{\alpha}}(\omega)\coth(2W(\omega)\mathbb{K})\right]W(\omega)\,e^{i\omega\tau}\,.    
\end{equation}
Therefore,
\begin{eqnarray}
-\left.\frac{\delta^{2} S_{\rm fluc}^{(\phi,\theta)}}
    {\delta A^{\rm L}_{\alpha} (\tau)
     \delta A^{\rm L}_{\alpha} (\tau')} \right|_{A=0} &\! = \! &  
     - \frac{\sqrt\lambda}{4\pi^2}
\int\limits_{-\infty}^{\infty}\!
d\omega\,
\coth(2W(\omega)\mathbb{K}) W(\omega)\,e^{i\omega(\tau-\tau')}\,,
\label{fourier1}
      \\
     -\left.\frac{\delta^{2} S_{\rm fluc}^{(\phi,\theta)}}{\delta A^{\rm R}_{\alpha} (\tau)
     \delta A^{\rm L}_{\alpha} (\tau')} \right|_{A=0} &\! = \! &  
     \frac{\sqrt\lambda}{4\pi^2} 
\int\limits_{-\infty}^{\infty}\!
d\omega\, 
\frac{W(\omega)}{{\sinh}(2W(\omega)\mathbb{K})}
\,e^{i\omega(\tau-\tau')}\,.
\label{fourier2}
\end{eqnarray}

A few comments are in order. First, the treatment of the two boundaries and the reduction of the equations of motion to those of a massive scalar field in flat space is similar to that of fluctuations on a world-sheet ending on two concentric circles \cite{Correa:2023uxp}. In that case, the results are expressed by a Fourier series rather than a Fourier transform.

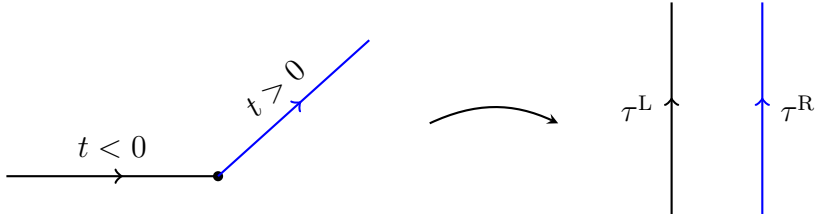
\begin{figure}
    \centering
\begin{tikzpicture}[
midarrow/.style={postaction={decorate,decoration={markings,mark=at position 0.55 with {\arrow[scale=1.2]{>}}}}},thick]
\draw[midarrow] (-2.8, 0) -- (0,0) node[midway, sloped, above=2.5pt] {$t < 0$};
\filldraw (0,0) circle (1.5pt);
\draw[midarrow,blue] (0,0) -- (2, 1.8) node[midway, sloped, above=2pt] {\color{black}$t > 0$};
\draw[->, >=stealth, thick] (2.8, 0.7) to[bend left=25] (4.5, 0.7);
\draw[midarrow] (6, -0.5) -- (6, 2.3) node[midway, left=2.5pt] {$\tau^{\rm L}$};
 \draw[midarrow,blue] (7.2, -0.5) -- (7.2, 2.3) 
        node[midway, right=2.5pt] {\color{black}$\tau^{\rm R}$};
\end{tikzpicture}
    \caption{Mapping a cusped line to two lines.}
    \label{fig:change}
\end{figure}

Second, we use global AdS coordinates in this dual description, so that the world-sheet ends in two parallel lines rather than in a cusped line. Thus, to match both descriptions, it is necessary to introduce the following change of coordinates 
\begin{equation} 
 t = e^{\kappa\, \tau^{\rm R}} \quad \text{for } t>0 \quad \text{and } \quad
 t = -e^{\kappa\, \tau^{\rm L} } \quad \text{for } t<0
 \quad \text{with } \kappa = b\sqrt{\frac{1-k^2}{1+b^2}}\,.
\end{equation}
This maps the cusped line to two infinite lines, as depicted in fig. \ref{fig:change}.

For example, the implementation of this change of coordinates for $t_1<t_2<0$ gives
\begin{eqnarray}
\frac{\langle\!\langle {\cal O}_{\alpha}(\tau_{1}^{\rm L}) {\cal O}_{\alpha}(\tau_{2}^{\rm L})
{\cal C}_\beta (-\infty) 
{\cal C}_\beta (\infty) 
\rangle\!\rangle }{\langle\!\langle
{\cal C}_\beta (-\infty) 
{\cal C}_\beta (\infty) 
\rangle\!\rangle }
&\!=\!& \frac{2B\, f(u)}{\left(e^{\kappa\,\tau_{1}^{\rm L}}-e^{\kappa\,\tau_{2}^{\rm L}}\right)^2}
\frac{dt_1}{d\tau_{1}^{\rm L}}  \frac{dt_2}{d\tau_{2}^{\rm L}}  
\nonumber\\
&\!=\!& \frac{2B\, \kappa^2\, f(u)}{2\cosh(\kappa\, \tau_{1}^{\rm L}-\kappa\,\tau_{2}^{\rm L})-2}
\,,
\end{eqnarray}
Finally, using that for $t_3=0$ and $t_4\to\infty$
$u = 1-\frac{t_2}{t_1} = 1- e^{\kappa(\tau_{2}^{\rm L}-\tau_{1}^{\rm L})}$, we can simply write
\begin{equation}
 \frac{\langle\!\langle {\cal O}_{\alpha}(\tau_{1}^{\rm L}) {\cal O}_{\alpha}(\tau_{2}^{\rm L})
{\cal C}_\beta (-\infty) 
{\cal C}_\beta (\infty) 
\rangle\!\rangle }{\langle\!\langle
{\cal C}_\beta (-\infty) 
{\cal C}_\beta (\infty) 
\rangle\!\rangle } = 
\frac{2B\, \kappa^2\, (1-u)\,   f(u)}{u^2}\,.
\label{TTFF}
\end{equation}

Consider now the other ordering, in which light and heavy operators are intercalated \eqref{2T2Fb}. Now we set $t_2=0$ and $t_4\to\infty$. Inserted at $t_1<0<t_3$, the local operators become at opposite lines in $\tau$ coordinates
\begin{eqnarray}
 \frac{\langle\!\langle {\cal O}_{\alpha}(\tau_{1}^{\rm L}) 
{\cal C}_\beta (-\infty) {\cal O}_{\alpha}(\tau_{3}^{\rm R})
{\cal C}_\beta (\infty) 
\rangle\!\rangle }{\langle\!\langle
{\cal C}_\beta (-\infty) 
{\cal C}_\beta (\infty) 
\rangle\!\rangle }
&\!=\!& -\frac{2B\, \tilde{f}(u)}{\left(e^{\kappa\,\tau_{1}^{\rm L}}+e^{\kappa\,\tau_{3}^{\rm R}}\right)^2}
\frac{dt_1}{d\tau_{1}^{\rm L}}  \frac{dt_3}{d\tau_{3}^{\rm R}}  
\nonumber\\
&\!=\!& \frac{2B\,\kappa^2\, \tilde{f}(u)}{2\cosh(\kappa\,\tau_{1}^{\rm L}-\kappa\,\tau_{3}^{\rm R})+2}
\,,
\end{eqnarray}
In this other case, $u = \frac{t_1}{t_1-t_3} = (1+ e^{\kappa(\tau_{3}^{\rm R}-\tau_{1}^{\rm L})})^{-1}$. Thus,
\begin{equation}
 \frac{\langle\!\langle {\cal O}_{\alpha}(\tau_{1}^{\rm L}) 
{\cal C}_\beta (-\infty) {\cal O}_{\alpha}(\tau_{3}^{\rm R})
{\cal C}_\beta (\infty) 
\rangle\!\rangle }{\langle\!\langle
{\cal C}_\beta (-\infty) 
{\cal C}_\beta (\infty) 
\rangle\!\rangle } =
2B\,
\kappa^2 u(1-u)\,   \tilde{f}(u)\,.
\label{TFTF}
\end{equation}

As a final observation,  notice that \eqref{fourier1} and \eqref{fourier2}  depend on $\tau$ and $\tau'$ through the difference in both cases. This implies that, using $\kappa(\tau_2^{\rm L}-\tau_1^{\rm L}) =\log(1-u)$ in one case and $\kappa(\tau_3^{\rm R}-\tau_1^{\rm L}) =\log(\frac{1-u}{u})$ in the other,  they are functions of the cross-ratio $u$ alone. This reinforces the interpretation of cusps as primary operators.

\subsection{Case $\theta=0$}
\label{geometriccusp}

Before turning to the most general analysis, in terms of an expansion in powers of $M$, we will consider the purely geometrical cusp with $\theta=0$. Notice that
\begin{equation}
 \theta^2 =  4 \mathbb{K}(k^2)^2 M^2\,.
\end{equation}

Thus, the purely geometrical cusp with $\theta=0$ requires  $M=0$, which is achieved by setting $b^2 = \frac{1}{k^2}-2$, and then $W(\omega)=|\omega|$. The value of the geometrical cusp is given by $k$,
\begin{equation}
 \phi = 2 k\sqrt{\frac{1-2k^2}{1-k^2}}
 \Pi(\tfrac{k^2}{1-k^2}|k^2)\,.
 \label{phitheta0}
\end{equation}
For $W(\omega)=|\omega|$, the inverse Fourier transforms \eqref{fourier1} and \eqref{fourier2} are easily done and we get, 
\begin{eqnarray}
 \!\! \frac{\langle\!\langle \mathbb{T}_a(\tau_{1}^{\rm L}) \mathbb{T}_a(\tau_{2}^{\rm L})
{\cal C}_3 (-\infty,\phi) 
{\cal C}_3 (\infty,-\phi) 
\rangle\!\rangle }{\langle\!\langle
{\cal C}_3 (-\infty,\phi) 
{\cal C}_3 (\infty,-\phi) 
\rangle\!\rangle } &\!=\!& 
\frac{\sqrt\lambda}{32\mathbb{K}(k^2)^2}\frac{1}{\sinh^2\left(\frac{\pi(\tau_2^{\rm L}-\tau_1^{\rm L})}{4\mathbb{K}(k^2)}\right)}\,\label{sinh},   
\\
 \!\! \frac{\langle\!\langle \mathbb{T}_a(\tau_{1}^{\rm L}) 
{\cal C}_3 (-\infty,\phi) 
\mathbb{T}_a(\tau_{3}^{\rm R})
{\cal C}_3 (\infty,-\phi) 
\rangle\!\rangle }{\langle\!\langle
{\cal C}_3 (-\infty,\phi) 
{\cal C}_3 (\infty,-\phi) 
\rangle\!\rangle } &\!=\!& 
\frac{\sqrt\lambda}{32\mathbb{K}(k^2)^2 }\frac{1}{\cosh^2\left(\frac{\pi(\tau_3^{\rm R}-\tau_1^{\rm L})}{4\mathbb{K}(k^2)}\right)}\,\label{cosh}.
\end{eqnarray}
These holographic results give the correlators \eqref{TTCC} and \eqref{TCTC} in the strong coupling limit. From them, and using that $\kappa=\sqrt{1-2k^2}$ in this case, we can infer the strong coupling limit of the corresponding cross-ratio functions
\begin{eqnarray}
   g(u,\phi)  &\!=\!& 
   \frac{\pi^2}{4\mathbb{K}^2(1-2k^2)}\frac{u^2}{1-u}\left((1-u)^{\frac{\pi}{4\mathbb{K}\sqrt{1-2k^2}}}-(1-u)^{-\frac{\pi}{4\mathbb{K}\sqrt{1-2k^2}}}\right)^{-2}
   \label{holographicg}
     \\
      \tilde{g} (u,\phi) &\!=\!& 
      \frac{\pi^2}{4\mathbb{K}^2(1-2k^2)}\frac{1}{u(1-u)}\left(\left(\tfrac{1-u}{u}\right)^{\frac{\pi}{4\mathbb{K}\sqrt{1-2k^2}}}+\left(\tfrac{1-u}{u}\right)^{-\frac{\pi}{4\mathbb{K}\sqrt{1-2k^2}}}\right)^{-2}
         \label{holographicgt}
\end{eqnarray}
These two functions of the cross-ratio
are expected to satisfy certain Bose symmetry relations, as a consequence of  the transposition of the two identical tilt operators. Indeed, 
\begin{align}
g(u,\phi) = &\  g\left(\tfrac{u}{u-1},\phi\right)\,,
\end{align}
reflects the symmetry under exchanging $t_1\leftrightarrow t_2$ in \eqref{2T2F}, while
\begin{align}
\tilde g(u,\phi) = &\ \tilde g(1-u,\phi)\,,
\end{align}
reflects the symmetry under exchanging $t_1\leftrightarrow t_3$ in \eqref{2T2Fb}.
These two Bose symmetry relations are an immediate consequence of the holographic description being invariant under $\tau\to-\tau$.

In addition to these Bose symmetry relations, one could wonder whether
the transposition $t_2\leftrightarrow t_3$ also implies a relation between $\tilde g$ and $g$. This is, however, not the case for \eqref{holographicg} and \eqref{holographicgt},
\begin{equation}
 \tilde g(u,\phi) \neq  g\left(\tfrac{1}{u},\phi\right)\,,
 \label{failure}
\end{equation}
(cf. \eqref{BoseFtF} for correlators of light local operators). 

The operator product expansion associated with the channel $u\to 1$, schematically reads
\begin{equation}
 \mathbb{T}(t){\cal C}(0)
 = \sum_h c_h^{\pm} |t|^{h-1-\Gamma_{\rm cusp}} {\cal O}_h(0)\,,
 \end{equation}
 where the $\pm$ in the coefficient indicates whether $\mathbb{T}$ approaches ${\cal C}$ from the right or from the left. The failure \eqref{failure} simply indicates that, for some of the exchanged operators, $c_h^+\neq c_h^-$.

\subsection{Generic case}
\label{generic}
In the previous subsection we studied the case where $M=0$, corresponding to a vanishing internal cusp $\theta=0$. In what follows, we will 
analyse the general case as an expansion in powers of $M$ or equivalently on $\theta$.
It turns out to be convenient to perform a series of rescalings: $w = 2 \mathbb{K}\, \omega$, $\tilde{\tau}=\frac{\tau}{2\mathbb{K}}$ and we also recall that
$\theta = 2\mathbb{K}M$. The integrals in \eqref{fourier1} and \eqref{fourier2} become
\begin{eqnarray}
-\left.\frac{\delta^{2} S_{\rm fluc}^{(\phi,\theta)}}
    {\delta A^{\rm L}_{\alpha} (\tau)
     \delta A^{\rm L}_{\alpha} (\tau')} \right|_{A=0} \!\! &\!\! = \!\! &  
     - \frac{\sqrt\lambda}{16\pi^2 \mathbb{K}^2}
\int\limits_{-\infty}^{\infty}\!
dw \,
\coth(\sqrt{w^2-\theta^2}) \sqrt{w^2-\theta^2}\,e^{iw(\tilde{\tau}-\tilde{\tau}')},
\label{fourier1t}
 \\
 -\left.\frac{\delta^{2} S_{\rm fluc}^{(\phi,\theta)}}{\delta A^{\rm R}_{\alpha} (\tau)
     \delta A^{\rm L}_{\alpha} (\tau')} \right|_{A=0} \!\! &\!\! = \!\! &  
     \frac{\sqrt\lambda}{16\pi^2 \mathbb{K}^2}
\int\limits_{-\infty}^{\infty}\!
d w\, 
\frac{\sqrt{w^2-\theta^2}}{{\sinh}(\sqrt{w^2-\theta^2})}
\,e^{i w(\tilde{\tau}-\tilde{\tau}')}\,.
\label{fourier2t}
\end{eqnarray}
In order to  perform the integrals, we make an expansion in powers of $\theta$. Notice that the Fourier transforms of both correlators are functions $h(\sqrt{w^2-\theta^2})$ and admit expansions of the form
\begin{equation}
\label{eq:expk^2}
h(\sqrt{w^2-\theta^2}) = h_0(w) + \theta^2 h_1(w)+ \theta^4  h_2(w)+ {\cal O}(\theta^6) \,,
\end{equation}
where 
\begin{equation}
h_n(w)=\frac{1}{n!}\frac{d^n}{d(\theta^2)^n} h(\sqrt{w^2-\theta^2})\big|_{\theta=0}    
\end{equation}
The fact that the dependence on $w$ and $\theta$ is through
$\sqrt{w^2-\theta^2}$, allows us to trade the derivatives with respect to $\theta^2$ by derivatives with respect to $w^2$. Then,
\begin{equation}
h_n(w)=\frac{1}{n!}\frac{d^n}{d(\theta^2)^n} h(\sqrt{w^2-\theta^2})\big|_{\theta=0}    
= \frac{(-1)^n}{n!}\frac{d^n }{d(w^2)^n} h_0(w)\,.
\label{hnsol}
\end{equation}
with $h(\sqrt{w^2-\theta^2}) \big|_{\theta=0}=h_0(w)$. The correlators \eqref{fourier1t} and \eqref{fourier2t} can be expanded in powers of $\theta$ as well
\begin{equation}
H(\tilde{\tau}-\tilde{\tau}',\theta)= H_0(\tilde{\tau}-\tilde{\tau}')+ \theta^2 H_1(\tilde{\tau}-\tilde{\tau}')+\theta^4 H_2(\tilde{\tau}-\tilde{\tau}')
+{\cal O} (\theta^6) 
\end{equation}
with $h_n(w)$ and  $H_n(\tilde{\tau})$ related by  Fourier transformation.

Notice that \eqref{hnsol} satisfies the recursive relation
\begin{equation}
 \frac{dh_{n-1}}{dw}=-2w n\,  h_{n}\,,
 \label{hnrecursion}
\end{equation}
from which it is straightforward to obtain a corresponding recursive relation for $H_n$,
\begin{equation}
\frac{dH_n}{d\tilde{\tau}}=-\frac{\tilde{\tau}}{2n} H_{n-1}\,.
 \label{Hnrecursion}
\end{equation}
Thus, instead of anti-transforming each $h_n$ to obtain $H_n$, we can anti-transform $h_0$, which is essentially what we did in subsection \ref{geometriccusp}, and use the relation \eqref{Hnrecursion} to obtain a generic $H_n$. Solving the integrals of this recursion and demanding that
$\lim_{\tilde{\tau}\rightarrow\pm\infty}H_n(\tilde{\tau})\rightarrow 0$, we obtain
\begin{align}
\label{eq:Fnsol1}        
H_n^{\rm LL}(\tilde\tau) = & \ \sum_{l=0}^{n} {C}_{n,l} 
|\tilde{\tau}|^{n-l}\text{Li}_{n+l-1}(+ e^{-\pi|\tilde{\tau}|})
\\
\label{eq:Fnsol2}        
H_n^{\rm RL}(\tilde\tau) = & \ \sum_{l=0}^{n} {C}_{n,l} 
|\tilde{\tau}|^{n-l}
\text{Li}_{n+l-1}(- e^{-\pi|\tilde{\tau}|})
\end{align} 
for $\eqref{fourier1t}$ and $\eqref{fourier2t}$. The coefficients in these expansions are
\begin{align}
C_{n,l} = &\ -\frac{\pi}{(2\pi)^{n+l-1}}\frac{(n+l)!}{ n! l! (n-l)!} 
\end{align}
and the full expressions for the integrals \eqref{fourier1t} and \eqref{fourier2t} are
\begin{eqnarray}
-\left.\frac{\delta^{2} S_{\rm fluc}^{(\phi,\theta)}}
    {\delta A^{\rm L}_{\alpha} (\tau)
     \delta A^{\rm L}_{\alpha} (\tau')} \right|_{A=0} &\!\! = \!\! &  
     -\frac{\sqrt\lambda}{16\pi^2 \mathbb{K}^2}
\sum_{n=0}^{\infty}\sum_{l=0}^{n} C_{n,l} 
\left(\tfrac{|\tau-\tau'|}{2\mathbb{K}}\right)^{n-l}
\text{Li}_{n-1+l}\left(x\right) \theta^{2n},
\label{fourier1sol}
      \\
     -\left.\frac{\delta^{2} S_{\rm fluc}^{(\phi,\theta)}}{\delta A^{\rm R}_{\alpha} (\tau)
     \delta A^{\rm L}_{\alpha} (\tau')} \right|_{A=0} &\! = \! &  
     \frac{\sqrt\lambda}{16\pi^2 \mathbb{K}^2}
     \sum_{n=0}^{\infty}\sum_{l=0}^{n} 
      C_{n,l} 
\left(\tfrac{|\tau-\tau'|}{2\mathbb{K}}\right)^{n-l}
\text{Li}_{n-1+l}\left(-x\right) \theta^{2n}
    \label{fourier2sol}
\end{eqnarray}
where $x= e^{-\frac{\pi}{2\mathbb{K}}|\tau-\tau'|}$. 

As presented, the solutions we have found are convenient for studying the case in which the internal cusp $\theta$ is small. In order to explore  other parametric limits it would be desirable to somehow sum the series expressions in \eqref{fourier1sol} and \eqref{fourier2sol}. We can progress in this direction  using the definition of the polylogarithms as  a series and interchanging the sums. First, we write :
\begin{equation}
 \text{Li}_{l+n-1}\left(\pm x \right)=\sum _{m=1}^{\infty } \frac{ \left(\pm x\right)^m}{m^{l+n-1}}   
\end{equation}
Then, we perform the finite sum in $l$. This finite sum  gives $K_{n+\frac{1}{2}}\left(m\pi \frac{|\tau-\tau'|}{2\mathbb{K}}\right)$  with additional factors. Then, we sum in $n$ using the Bessel multiplication formula,
\begin{equation}
\mu^{-\nu} K_{\nu}(\mu z) =\sum_{n=0}^{\infty} \frac{(-1)^n}{n!} \left[ \frac{(\mu^2 - 1)z}{2} \right]^n K_{\nu+n}(z),
\end{equation}
where we have for this case $z=m\pi \frac{|\tau-\tau'|}{2\mathbb{K}}$ and $\mu=\tfrac{1}{\pi m}\sqrt{m^2 \pi^2-\theta^2}$, so we are just left with a sum on $m$ $K_{\frac{1}{2}}\left(\sqrt{m^2\pi^2-\theta^2}\frac{|\tau-\tau'|}{2\mathbb{K}}\right)$, replacing  the value of $K_{1/2}$ and putting all factors together we get,
\begin{eqnarray}
-\left.\frac{\delta^{2} S_{\rm fluc}^{(\phi,\theta)}}
    {\delta A^{\rm L}_{\alpha} (\tau)
     \delta A^{\rm L}_{\alpha} (\tau')} \right|_{A=0} &\!\! = \!\! &  
    \frac{\sqrt\lambda}{8\pi\mathbb{K}^2} \sum_{m=1}^{\infty}   \frac{\pi^2 m^2 }{ \sqrt{\pi ^2 m^2-\theta^2}}e^{-\frac{|\tau-\tau'|}{2\mathbb{K}} \sqrt{\pi ^2 m^2-\theta^2}},
\label{fourier1solt}
      \\
     -\left.\frac{\delta^{2} S_{\rm fluc}^{(\phi,\theta)}}{\delta A^{\rm R}_{\alpha} (\tau)
     \delta A^{\rm L}_{\alpha} (\tau')} \right|_{A=0} &\! = \! &  \frac{\sqrt\lambda}{8\pi\mathbb{K}^2} \sum_{m=1}^{\infty}   \frac{(-1)^{m+1} \pi^2 m^2 }{ \sqrt{\pi ^2 m^2-\theta^2}}e^{-\frac{|\tau-\tau'|}{2\mathbb{K}} \sqrt{\pi ^2 m^2-\theta^2}}.
\label{fourier2solt}
\end{eqnarray}
While this new formula is also an infinite series, it is valid for any value of $\theta$ and $\phi$.

Because, we will use it in the next section, we conclude this one by writing down our solutions in the cases (i) $\phi=0$ and $\theta$ small and (ii) $\theta=0$ and $\phi$ small.

The case $\phi=0$ is obtained by taking $b\to\infty$ in the generic case description. If we then expand for small $k$, we see that
\begin{equation}
    \theta = i\pi k + {\cal O}(k^3)
    \quad \Rightarrow \quad k^2 = -\frac{\theta^2}{\pi^2}+{\cal O}(\theta^4)\,.
\end{equation}
We use this and keep only $n=0,1$ in the sum \eqref{fourier1sol}. In this limit,
the factor between global and world-sheet times becomes $\kappa=\sqrt{1-k^2}$ and 
we obtain
\begin{align}
 \frac{1}{\kappa^2}\frac{\langle\!\langle \mathbb{T}_1(\tau_{1}^{\rm L}) \mathbb{T}_1(\tau_{2}^{\rm L})
{\cal F}_5 (-\infty,\theta) 
{\cal F}_5 (\infty,-\theta) 
\rangle\!\rangle }{\langle\!\langle
{\cal F}_5 (-\infty,\theta) 
{\cal F}_5 (\infty,-\theta) 
\rangle\!\rangle } = & \frac{\sqrt\lambda}{2\pi^2}\frac{(1-u)}{u^2}
\label{holoresultf}
 \\
& \hspace{-2.8cm}
\times
\underbrace{\left(1 -\frac{\theta^2}{2\pi^2}\left(1 + 
\left(\frac{1}{u} - \frac{1}{2} +  u\right)\log(1-u) 
 + \frac{u^2}{1-u}\log u\right)\right)}_{f(u,\theta)}
  \nonumber
\,.
\end{align}

If one of the tilts is intercalated between the two DCOs, we get
\begin{align}
 \frac{1}{\kappa^2} \frac{\langle\!\langle \mathbb{T}_1(\tau_{1}^{\rm L}) 
{\cal F}_5 (-\infty,\theta) 
\mathbb{T}_1(\tau_{2}^{\rm L})
{\cal F}_5 (\infty,-\theta) 
\rangle\!\rangle }{\langle\!\langle
{\cal F}_5 (-\infty,\theta) 
{\cal F}_5 (\infty,-\theta) 
\rangle\!\rangle } = & \frac{\sqrt\lambda}{2\pi^2} u(1-u)
\label{holoresultf2}
 \\
& \hspace{-4.4cm}
\times
\underbrace{\left(1 -\frac{\theta^2}{2\pi^2}\left(1 + 
\left(\frac{1}{u} - \frac{1}{2} +  u\right)\log(1-u) +
   \left(\frac12 + \frac1{1 - u} - u\right)\log u\right)\right)}_{\tilde{f}(u,\theta)}
  \nonumber
\,.
\end{align}

The case $\theta=0$ was analysed separately in \ref{geometriccusp}. For the expansion of
\eqref{holographicg} and \eqref{holographicgt} in powers of $\phi$, we use
\begin{equation}
k =\frac{\phi}{\pi}-\frac{1}{4}\frac{\phi^3}{\pi^3}  +{\cal O}(\phi^5)\,,
\end{equation}
and we obtain
\begin{align}
g(u,\phi) =\, & 1+\frac{\phi^2}{\pi^2}\left(\frac32+\frac{3(2-u)}{4u}\log(1-u)\right)  +{\cal O}(\phi^4)  \,,
\label{gsmallphi}
\\
\tilde{g}(u,\phi) =\, & 1+\frac{\phi^2}{\pi^2}\left(\frac32-\frac{3(1-2u)}{4}\log(1-u)+\frac{3(1-2u)}{4}\log u\right)  +{\cal O}(\phi^4)  \,,
\label{gtsmallphi}
\end{align}

\section{Small cusp angles}
\label{smallgeometriccusp}

We have been referring to the cusp operators as either DCOs or heavy operators interchangeably, because their scaling dimensions are of order  $\sqrt\lambda$ in the strong coupling limit and they back-react on the world-sheet geometry. However, as these scaling dimensions also depend on the values of the cusp angles, we can make them as small as we want by taking $\theta$ and $\phi$ sufficiently small. In other words, by dialling the values of the cusp angles, the character of the DCOs interpolates between being heavy or light. Considering correlators involving DCOs with small cusp angles turns out to be interesting as it gives rise to relations that integrated correlation functions of light operators must fulfil. 

For instance, the correlators between two tilts and two internal cusp DCOs \eqref{2T2F} admit an expansion in even powers of the cusp angle using
\begin{eqnarray}
    f(u,\theta) &\!=\!& f_0(u) +\theta^2 f_2(u) 
    +\theta^4 f_4(u) +
    {\cal O}(\theta^6)
    \\
    \Gamma_{\rm cusp}(\theta) &\!=\!& \theta^2 B + {\cal O}(\theta^4)
    \\
    {\cal N}_{\cal F}(\theta) &\!=\!& 1 + \theta^2 n_2 + {\cal O}(\theta^4)
\end{eqnarray}
Thus, setting $t_3=0$ and $t_4=L\to\infty$, the small cusp angle expansion of \eqref{2T2F} becomes
\begin{equation}
\!\!    \langle\!\langle \mathbb{T}_1(t_1) \mathbb{T}_1(t_2)
{\cal F}_5 (0,\theta) 
{\cal F}_5 (\infty,-\theta) 
\rangle\!\rangle  =  \frac{2B\, f_0}{t_{12}^2}  +
  \theta^2 \frac{2B}{t_{12}^2} \left(f_0 n_2 +f_2-2B f_0\log(\tfrac{L}{\epsilon})\right)+ {\cal O}(\theta^4)\,,
\label{2T2Fexpanded}
\end{equation}
We can proceed as in \eqref{2Fssapp0} and express
the expansion of \eqref{2T2F} in terms of integrals of correlators of light operators alone
\begin{eqnarray}
\label{2T2Fss}
\!\langle\!\langle \mathbb{T}_1(t_1) \mathbb{T}_1(t_2)
{\cal F}_5 (0,\theta) 
{\cal F}_5 (\infty,-\theta) 
\rangle\!\rangle 
&\!\! = \!\! &  \langle\!\langle 
\mathbb{T}_1(t_1) \mathbb{T}_1(t_2) \rangle\!\rangle
+ \theta \int_{0}^{\infty}\!dt\, \langle\!\langle 
\mathbb{T}_1(t_1) \mathbb{T}_1(t_2)
\mathbb{T}_5 (t) \rangle\!\rangle\\
&& \hspace{-6.5cm}
  -\frac12 \theta^2 \int_{0}^{\infty}\!dt\, \langle\!\langle \mathbb{T}_1(t_1) \mathbb{T}_1(t_2)\Phi_6(t) \rangle\!\rangle  
   + \frac{1}{2}\theta^2 \int_{0}^{\infty}\!dt\, \int_{0}^{\infty}\!dt'\, 
    \langle\!\langle \mathbb{T}_1(t_1) \mathbb{T}_1(t_2)
\mathbb{T}_5 (t) \mathbb{T}_5 (t') \rangle\!\rangle 
+{\cal O}(\theta^3)\,. \nonumber
 \end{eqnarray}
The comparison of \eqref{2T2Fexpanded} with \eqref{2T2Fss} yields interesting relations. The first one, that $f_0(u)=1$ is rather obvious because of the vanishing cusp ${\cal F}_a (t,0) \equiv\mathds{1}$. The vanishing of the order $\theta$ term in \eqref{2T2Fss} is also immediate, as there is no possible way to write 3 $SO(5)$ indices in terms of the $SO(5)$ invariant tensors. Finally, the comparison of the $\theta^2$ terms does provide a non-trivial relation,
\begin{eqnarray}
    \frac{2B}{t_{12}^2} \left(n_2+f_2(u)-2B \log(\tfrac{L}{\epsilon})\right) &\!\! = \!\! & 
   -\frac12 \int_{0}^{\infty}\!dt\, \langle\!\langle \mathbb{T}_1(t_1) \mathbb{T}_1(t_2)\Phi_6(t) \rangle\!\rangle \nonumber\\ 
   &&+ \frac{1}{2}\int_{0}^{\infty}\!dt\, \int_{0}^{\infty}\!dt'\, 
    \langle\!\langle \mathbb{T}_1(t_1) \mathbb{T}_1(t_2)
\mathbb{T}_5 (t) \mathbb{T}_5 (t') \rangle\!\rangle
\label{condi1}
\end{eqnarray}
The Witten diagrams computing the correlators in the r.h.s. of \eqref{condi1} can be split into disconnected and connected ones. Since,
\begin{eqnarray}
 \frac{2B}{t_{12}^2} \left(n_2-2B \log(\tfrac{L}{\epsilon})\right) &\!\! = \!\! & 
   \langle\!\langle \mathbb{T}_1(t_1) \mathbb{T}_1(t_2) \rangle\!\rangle
    \left(- \frac{1}{2} \!\!\int_{0}^{\infty}\!\!dt \, 
    \langle\!\langle 
\Phi_6 (t) \rangle\!\rangle 
   + \frac{1}{2} \!\!\int_{0}^{\infty}\!\!dt\int_{0}^{\infty}\!dt'\,
    \langle\!\langle 
\Phi_5 (t) \Phi_5 (t') \rangle\!\rangle \right)
\nonumber  
 \end{eqnarray}
we are left with
\begin{eqnarray}
    \frac{2B}{t_{12}^2} f_2(u) &\!\! = \!\! & 
   -\frac12 \int_{0}^{\infty}\!dt\, \langle\!\langle \mathbb{T}_1(t_1) \mathbb{T}_1(t_2)\Phi_6(t) \rangle\!\rangle_{\rm c} 
   \nonumber\\ && 
   + \frac{1}{2}\int_{0}^{\infty}\!dt\, \int_{0}^{\infty}\!dt'\, 
    \langle\!\langle \mathbb{T}_1(t_1) \mathbb{T}_1(t_2)
\mathbb{T}_5 (t) \mathbb{T}_5 (t') \rangle\!\rangle_{\rm c}\,,
\label{condi1conn}
\end{eqnarray}
where the subindex ``c'' stands for connected. We should also  emphasize at this point that the condition \eqref{condi1conn} is meant to be valid to all orders in the coupling constant dependence. We will test this relation in the strong coupling limit.
The two-particle decomposition of $\Phi_6$ \eqref{Phi6proposal} allows to write the r.h.s. of \eqref{condi1conn} in terms of two 4-point correlators, which will be evaluated in the strong coupling limit using the expressions given in the appendix  \ref{app:4pt}.

The strong coupling leading order of the lhs of \eqref{condi1conn} is directly read from the holographic result \eqref{holoresultf},
\begin{equation}
   \frac{2B}{t_{12}^2} f_2(u)  =
-\frac{\sqrt\lambda}{4\pi^4}
\frac{1}{t_{12}^2}\left(1 + 
\left(\frac{1}{u} - \frac{1}{2} + u\right)\log(1-u) 
  +\frac{u^2}{1-u}\log u\right)
      + {\cal O}(\lambda^0)
      \label{lhscondconn}
\end{equation}
The first of the two integrals in the r.h.s. of
\eqref{condi1conn} is 
\begin{eqnarray}
     I_1 &\! =\!&  - \frac{\epsilon}5 \sum_{a=1}^5 
     \int_{0}^{\infty}\!dt\, \langle\!\langle \mathbb{T}_1(t_1) \mathbb{T}_1(t_2)
     \mathbb{T}_a(t-\tfrac{\epsilon}{2})\mathbb{T}_a(t+\tfrac{\epsilon}{2})\rangle\!\rangle_{\rm c} 
    \nonumber 
     \\
&\! =\!& 
 -\frac{\gamma^{2}}{t_{12}^{2}\,\epsilon}\frac{1}{\sqrt{\lambda}}\int_{0}^{\infty}\!dt\, T_S^{(1)}(u_\epsilon)    
 \label{fisrtinte}
\end{eqnarray}
with
\begin{equation}
    u_\epsilon =\frac{(t_2-t_1)\epsilon}{(t-t_1-\tfrac{\epsilon}{2})(t-t_2+\tfrac{\epsilon}{2})}
\end{equation}
Using \eqref{TS1}, it is straightforward to show that $I_1$ goes to 0 as $\epsilon$ goes to 0.

We now turn to the second integral in the r.h.s. of
\eqref{condi1conn}
\begin{eqnarray}
   I_2 &\! =\!&  \frac{1}{2}\int_{0}^{\infty}\!\!\!dt\int_{0}^{\infty}\!\!\!dt'
    \langle\!\langle \mathbb{T}_1(t_1) \mathbb{T}_1(t_2)
\mathbb{T}_5 (t) \mathbb{T}_5 (t') \rangle\!\rangle_{\rm c}
\\
&\! = \! & 
\frac{\sqrt\lambda}{4\pi^4} \int_{0}^{\infty}\!\!\!dt \int_{t}^{\infty}\!\!\!dt'\,
\frac{T_S^{(1)}(v)-\frac{2}{5}T_T^{(1)}(v)}{t_{12}^2(t-t')^2}
\end{eqnarray}
where now
\begin{equation}
    v =\frac{(t_2-t_1)(t'-t)}{(t-t_1)(t'-t_2)}\,.
    \label{defv}
\end{equation}
To evaluate this integral, we use \eqref{TS1} and \eqref{TT1} and  change coordinates from $(t,t')$ to $(v,x)$, with $v$ defined in \eqref{defv} and,
\begin{equation}
    x  = \frac{(t_2-t_1)}{(t-t_1)}\,.
\end{equation}    
In these new coordinates $v<x$ and $0<x<u = 1-t_2/t_1$. The integral remains free of UV and IR divergences. By explicit computation we obtain
\begin{align}
I_2 = & \,  \frac{\sqrt\lambda}{4\pi^4}\frac{1}{t_{12}^2} \int_{0}^{u}\! \frac{dx}{x-1} \int_{0}^{x}\!dv\left(
\tfrac{2 - 2 v + v^2}{(1 - v)^2 v^2}
(1-v + v^2\log v) + \tfrac{(1-v) (2 + v +v^2)}{v^3}
 \log(1 - v)\right)\nonumber
 \\
 = & -\frac{\sqrt\lambda}{4\pi^4}
\frac{1}{t_{12}^2}\left(1 + 
\left(\frac{1}{u} - \frac{1}{2} +  u\right)\log(1-u) 
  + \frac{u^2}{1-u}\log u\right)
      + {\cal O}(\lambda^0)\,,
\end{align}
which exactly matches \eqref{lhscondconn} and therefore verifies the condition \eqref{condi1conn} in the strong coupling limit.

We can repeat the same analysis for the correlator \eqref{TTCC} between two tilts and two geometrical cusp DCOs, using in this case 
\begin{eqnarray}
    g(u,\phi) &\!=\!& 1 +\phi^2 g_2(u) + {\cal O}(\phi^4)
    \\
    \Gamma_{\rm cusp}(\phi) &\!=\!& -\phi^2 B + {\cal O}(\phi^4)
    \\
    {\cal N}_{\cal C}(\phi) &\!=\!& 1 + \phi^2 p_2 + {\cal O}(\phi^4)
\end{eqnarray}
where we have already set $g_0=1$. So, the small cusp angle expansion is
\begin{equation}
\!\!    \langle\!\langle \mathbb{T}_1(t_1) \mathbb{T}_1(t_2)
{\cal C}_1 (0,\phi) 
{\cal C}_1 (\infty,-\phi) 
\rangle\!\rangle  =  \frac{2B}{t_{12}^2}  +
  \phi^2 \frac{2B}{t_{12}^2} \left(p_2 +g_2+2B \log(\tfrac{L}{\epsilon})\right)+ {\cal O}(\phi^4)\,,
\label{2T2Cexpanded}
\end{equation}
Also for \eqref{TTCC} we write an expansion analogous to the one in \eqref{2Css}, 
\begin{eqnarray}
\!\langle\!\langle \mathbb{T}_1(t_1) \mathbb{T}_1(t_2)
{\cal C}_1 (0,\phi) 
{\cal C}_1 (\infty,-\phi) 
\rangle\!\rangle 
&\!\! = \!\! &  \langle\!\langle 
\mathbb{T}_1(t_1) \mathbb{T}_1(t_2) \rangle\!\rangle
+ \phi \int_{0}^{\infty}\!dt\, \langle\!\langle 
\mathbb{T}_1(t_1) \mathbb{T}_1(t_2)
\mathbb{D}_1 (t) \rangle\!\rangle t
\nonumber
\\
&& \hspace{-4cm}
  +\frac12 \phi^2 \int_{0}^{\infty}\!dt\, \langle\!\langle \mathbb{T}_1(t_1) \mathbb{T}_1(t_2)\Phi_6(t) \rangle\!\rangle  
  +\frac12 \phi^2 \int_{0}^{\infty}\!dt\, \langle\!\langle \mathbb{T}_1(t_1) \mathbb{T}_1(t_2)D_1\mathbb{D}_1(t) \rangle\!\rangle t^2 
  \nonumber
  \\
&& \hspace{-4cm}
   + \frac{1}{2}\phi^2 \int_{0}^{\infty}\!dt\, \int_{0}^{\infty}\!dt'\, 
    \langle\!\langle \mathbb{T}_1(t_1) \mathbb{T}_1(t_2)
\mathbb{D}_1 (t) \mathbb{D}_1 (t') \rangle\!\rangle t\, t' +{\cal O}(\phi^3)\,. 
 \end{eqnarray}
Since the correlator in the ${\cal O}(\phi)$ integral is vanishing,
\begin{eqnarray}
\frac{2B}{t_{12}^2} \left(p_2 +g_2+2B \log(\tfrac{L}{\epsilon})\right) \! &\!\! = \!\! & \!
  \frac12 \int_{0}^{\infty}\!\!\!dt \langle\!\langle \mathbb{T}_1(t_1) \mathbb{T}_1(t_2)\Phi_6(t) \rangle\!\rangle  
  +\frac12  \int_{0}^{\infty}\!\!\!dt \langle\!\langle \mathbb{T}_1(t_1) \mathbb{T}_1(t_2)D_1\mathbb{D}_1(t) \rangle\!\rangle t^2 
  \nonumber
  \\
&& \hspace{-0.6cm}
   + \frac{1}{2} \int_{0}^{\infty}\!dt\, \int_{0}^{\infty}\!dt'\, 
    \langle\!\langle \mathbb{T}_1(t_1) \mathbb{T}_1(t_2)
\mathbb{D}_1 (t) \mathbb{D}_1 (t') \rangle\!\rangle t\, t' \,. 
\label{condi2}
 \end{eqnarray}
As before, we can split contributions to the integrals in the r.h.s. of \eqref{condi2} into disconnected and connected ones. Given that
\begin{eqnarray}
\frac{2B}{t_{12}^2} \left(p_2 +2B \log(\tfrac{L}{\epsilon})\right) 
\! &\!\! = \!\! & \!
\langle\!\langle \mathbb{T}_1(t_1) \mathbb{T}_1(t_2) \rangle\!\rangle
    \left( \frac{1}{2} \!\int_{0}^{\infty}\!\!dt \, 
    \langle\!\langle 
\Phi_6 (t) \rangle\!\rangle 
+\frac{1}{2} \!\int_{0}^{\infty}\!\!dt \, 
    \langle\!\langle 
D_1 \mathbb{D}_1  (t) \rangle\!\rangle t^2 \right.
\nonumber
\\
&& \hspace{3.1cm}
   + \left.\frac{1}{2} \int_{0}^{\infty}\!dt\, \int_{0}^{\infty}\!dt'\, 
    \langle\!\langle
\mathbb{D}_1 (t) \mathbb{D}_1 (t') \rangle\!\rangle t\, t' \right)\,,
 \end{eqnarray}
the function $g_2(u)$ is expressed in terms of the connected correlators. Taking into account
that the integral of $\langle\!\langle \mathbb{T}_1(t_1) \mathbb{T}_1(t_2)\Phi_6(t) \rangle\!\rangle_{\rm c}$ vanishes for $\epsilon\to 0$, we are left with
\begin{eqnarray}
\frac{2B}{t_{12}^2}  g_2  \! &\!\! = \!\! & \!
  \frac12  \int_{0}^{\infty}\!\!\!dt \langle\!\langle \mathbb{T}_1(t_1) \mathbb{T}_1(t_2)D_1\mathbb{D}_1(t) \rangle\!\rangle_{\rm c} t^2 
  \nonumber
 \\
&& \hspace{0.6cm}
   + \frac{1}{2} \int_{0}^{\infty}\!dt\, \int_{0}^{\infty}\!dt'\, 
    \langle\!\langle \mathbb{T}_1(t_1) \mathbb{T}_1(t_2)
\mathbb{D}_1 (t) \mathbb{D}_1 (t') \rangle\!\rangle_{\rm c} t\, t' \,. 
\label{condi2conn}
 \end{eqnarray}
The relation \eqref{condi2conn} is also valid to all orders in the coupling constant expansion. We will test it to leading order in the strong coupling limit.

The strong coupling leading order of the l.h.s. of \eqref{condi2conn} is obtained 
from  \eqref{gsmallphi},
\begin{equation}
   \frac{2B}{t_{12}^2}  g_2
   =
   \frac{3}{4}\frac{\sqrt{\lambda}}{\pi^4}\frac{1}{t_{12}^2}
    \left(1-\big(\tfrac{1}{2}-\tfrac{1}{u}\big)\log(1-u)\right)
    + {\cal O}(\lambda^0)
      \label{lhscond2conn}
\end{equation}

The analysis of the r.h.s of \eqref{condi2conn}  is more subtle than in the previous case, as both integrals present UV divergences. It is only after summing them that we obtain a finite result. The first integral in the r.h.s of \eqref{condi2conn} is
\begin{eqnarray}
     I_3 &\! =\!&   
    \frac{\eta\, \epsilon}{2} \int_{0}^{\infty}\!dt\, \langle\!\langle \mathbb{T}_1(t_1) \mathbb{T}_1(t_2)
      \mathbb{D}_1(t-\tfrac{\epsilon}{2})\mathbb{D}_1(t+\tfrac{\epsilon}{2})
     \rangle\!\rangle_{\rm c} t^2
    \nonumber 
     \\
&\! =\!& 
 \frac{3\sqrt\lambda}{4\pi^4}\frac{\eta}{t_{12}^{2}\,\epsilon^3}
 \int_{0}^{\infty}\!dt\, F^{(1)}(u_\epsilon) t^2
 \nonumber 
     \\
&\! =\!& 
\frac{\sqrt\lambda}{2\pi^4}\frac{\eta}{t_{12}^{2}}\frac{t_1}{\epsilon}
 \left(\tfrac{u}{2}-1+\left(1-\tfrac{1}{u}\right)\log(1-u)\right)
 +{\cal O}(\epsilon)
 \label{3rdinte}
\end{eqnarray}
The double integral in the r.h.s of \eqref{condi2conn} is
\begin{eqnarray}
   I_4 &\! =\!&  \frac{1}{2}\int_{0}^{\infty}\!\!\!dt\int_{0}^{\infty}\!\!\!dt'
    \langle\!\langle \mathbb{T}_1(t_1) \mathbb{T}_1(t_2)
\mathbb{D}_1 (t) \mathbb{D}_1 (t') \rangle\!\rangle_{\rm c} t\,t'
\nonumber
\\
&\! = \! & 
\frac32\frac{\sqrt\lambda}{\pi^4} \int_{0}^{\infty}\!\!\!dt \int_{t}^{\infty}\!\!\!dt'\,
\frac{F^{(1)}(v)}{t_{12}^2(t-t')^4}t\,t'
\label{4thinte}
\\
&\! = \! & 
\frac{\sqrt\lambda}{\pi^4t_{12}^{2}}\frac{t_1}{\epsilon}
 \left(\tfrac{u}{2}-1+\left(1-\tfrac{1}{u}\right)\log(1-u)\right)
 +
 \frac{3}{4}\frac{\sqrt{\lambda}}{\pi^4}\frac{1}{t_{12}^2}
    \left(1-\big(\tfrac{1}{2}-\tfrac{1}{u}\big)\log(1-u)\right)
  +{\cal O}(\epsilon)\,.
 \nonumber
\end{eqnarray}
For $\eta=-2$, the UV divergences in $I_3+I_4$ cancel each other and the remainder exactly matches \eqref{lhscond2conn}. Therefore, the condition \eqref{condi2conn} is also verified in the strong coupling limit.

Similar identities apply to correlators where light operators and DCOs are intercalated. These are obtained by replacing $f_2$ with $\tilde f_2$ and $g_2$ with $\tilde g_2$ in \eqref{condi1conn} and \eqref{condi2conn}, while taking $t_2>0$. Their verification at strong coupling proceeds analogously.

By going to higher orders in the small cusp expansion, one can derive similar relations for integrated correlators involving more light operators. 
For example,
\begin{align}
\frac{2B}{t_{12}^2}\,f_4(u)
\;=\; &\;
\frac{1}{24}\int_0^\infty\!\! ds_1\;\dbra\T_1\T_1\,\Phi_6(s_1)\dket_{\rm c}\ 
 - \frac16\int_0^\infty\!\! ds_1\!\int_0^\infty\!\! ds_2\;
\dbra\T_1\T_1\,\T_5(s_1)\T_5(s_2)\dket_{\rm c}
\nonumber\\
+&\;\frac18\int_0^\infty\!\! ds_1\!\int_0^\infty\!\! ds_2\;
\dbra\T_1\T_1\,\Phi_6(s_1)\Phi_6(s_2)\dket_{\rm c}
\nonumber\\
-&\;\frac14\int_0^\infty\!\! ds_1\!\int_0^\infty\!\! ds_2\!\int_0^\infty\!\! ds_3\;
\dbra\T_1\T_1\,\T_5(s_1)\T_5(s_2)\Phi_6(s_3)\dket_{\rm c}
\nonumber\\
+&\;\frac{1}{24}\int_0^\infty\!\! ds_1\!\! \int_0^\infty\!\! ds_2\!\!\int_0^\infty\!\! ds_3\!\! \int_0^\infty\!\! ds_4\;
\dbra\T_1\T_1\,\T_5(s_1)\T_5(s_2)\T_5(s_3)\T_5(s_4)\dket_{\rm c}
\label{condi3conn}
\end{align}
where $\T_1\T_1$ stands for $\T_1(t_1)\T_1(t_2)$. Like \eqref{condi1conn} and \eqref{condi2conn}, this relation holds at any value of the coupling.

\section{Discussion}
\label{discu}
The main result of this paper is a holographic prescription for correlation functions of two heavy defect changing operators (DCOs) and an arbitrary number of light protected operators on the $1/2$ BPS Wilson line of ${\cal{N}}=4$ SYM. 
The introduction of a cusp on the Wilson line, with geometrical and internal cusp angles $(\phi,\theta)$, translates into a back-reaction on the dual worldsheet, which is no longer $AdS_2$. In this setup, the light operators are described by fluctuations around the new back-reacted geometry, and the normalized correlator is obtained
from functional derivatives of the on-shell action with respect to the boundary values of the fluctuations \eqref{proposal}. For two light local operators, it is sufficient to work with the quadratic action of fluctuations, and the result is given by the boundary-to-boundary propagators \eqref{fourier1solt} and
\eqref{fourier2solt}, obtained from their Fourier
space expressions \eqref{fourier1} and \eqref{fourier2} or directly from the bulk-to-bulk propagator (as shown in Appendix \ref{app green}).
We have considered both possible orderings ---HHLL and HLHL--- and verified that, up to a kinematical factor they only depend on the unique cross-ratio $u$. This validates the interpretation of cusps as primary operators of dimension $\Gamma_{\rm cusp}$.

For the purely geometric cusp $(\theta=0)$,
the Fourier expressions \eqref{fourier1} and \eqref{fourier2} are easily inverse-transformed to a closed form, yielding simple expressions for the functions $g(u,\phi)$ and $\tilde{g}(u,\phi)$ in \eqref{holographicg} and \eqref{holographicgt}. We have verified they satisfy the Bose symmetries associated with the exchange of the two tilts, $g(u,\phi) = g(\tfrac{u}{u-1},\phi)$ and $\tilde g(u,\phi) =  \tilde g(1-u,\phi)$. On the other hand, the fact that
$\tilde g(u,\phi) \neq  g(\tfrac{1}{u},\phi)$ 
reveals that moving the tilt towards the DCO from the left is not equivalent to doing so from the right.

For generic cusp angles, we provided two complementary series representations for the HHLL and HLHL correlators. The expansions \eqref{fourier1sol} and \eqref{fourier2sol} in powers of $\theta$ are well suited for the small internal cusp limit, while the expansions \eqref{fourier1solt} and \eqref{fourier2solt} are useful for exploring other parametric limits.

As a non-trivial consistency check, we considered the regime where the cusp angles are small.  In this limit, the HHLL and HLHL correlators must reduce to integrated correlators of light operators alone. This gives the relations \eqref{condi1conn} and \eqref{condi2conn}, which are valid at any value of the coupling, as they are derived from the definition of the DCOs. We verified these relations at strong coupling for both the internal and geometrical cusps, using the four-point functions of \cite{Giombi:2017cqn}. This supports the prescription \eqref{proposal}.

There are several directions in which this analysis can be extended. One is to increase the number of light insertions. For  $n-2$ light operators, one should consider the fluctuation action at $(n-2)^{\rm th}$ order. For example, for four light operators, one needs the quartic vertices to evaluate genuine contact Witten diagrams on the back-reacted worldsheet, using the bulk-to-boundary propagators that follow from the Green's functions of Appendix \ref{app green}. This would give access to HHLLLL correlators, which, in the small cusp angle limit, would also imply other relations for integrated correlators of six light operators. It would also be interesting to use our results to explore specific parametric regimes of the cusped configuration. In particular, in the ladder limit $i\theta\to \infty$ with $\lambda e^{i\theta}$ held fixed, the perturbative expansion is dominated by ladder diagrams and becomes exactly summable in certain cases \cite{Correa:2012nk,Cavaglia:2018lxi}. One could also explore our results in the fusion limit $\phi\to\pi$ or the supersymmetric limit $\phi =\pm \theta$, where $\Gamma_{\rm cusp}$ vanishes identically and the DCOs become BPS. In this latter case, the cusp insertions cease to be heavy no matter what values the angles take.

Finally, one could also generalize this to other field insertions from the displacement multiplet, as well as consider defect changing operators carrying a large R-charge, by allowing insertions of $Z^L$ at the tip of the cusps, whose dual classical worldsheet configuration is also known \cite{Giombi:2021zfb,Giombi:2022anm,Bonomi:2025ctm}.

\subsection*{Acknowledgements}

We would like to thank Jerem\'ias Aguilera-Damia, Nadav Drukker, Mart\'in Lagares, Ignacio Salazar Landea and Guillermo Silva for useful discussions. This work was partially supported by  PICT 2020-03749, PIP 02229, UNLP X1044. VIGR would like to acknowledge support from the ICTP through the Associates Programme. MC and NAI are supported by fellowships from CONICET (Argentina). AI models were used to check calculations as well as to edit the text. The ideas and content of this paper came entirely from the authors.

\appendix

\section{Four-point functions of light operators}
\label{app:4pt}

Here, we review the structure of the four-point functions of the protected light
insertions on the $1/2$-BPS Wilson line, namely the tilt operators $\T_a$ ($a=1,\dots,5$) and the displacement operators $\D_i$ ($i=1,2,3$), whose two-point functions were given in \eqref{2ptfunctions}. All four-point  functions depend on the unique independent cross-ratio
\begin{equation}
u=\frac{t_{12}t_{34}}{t_{13}t_{24}}\,,
\end{equation}
which satisfies $0<u<1$ for $t_1<t_2<t_3<t_4$.

In the case of four tilt operators, the $SO(5)$ tensor structure decomposes
into singlet, symmetric traceless and antisymmetric parts,
\begin{align}
\dbra \T_{a}(t_1)\T_{b}(t_2)\T_{c}(t_3)\T_{d}(t_4)\dket
&=\frac{\gamma^{2}}{t_{12}^{2}\,t_{34}^{2}}
\left(
\delta_{ab}\delta_{cd}\,T_S(u)
+(\delta_{ac}\delta_{bd}+\delta_{ad}\delta_{bc}
-\tfrac{2}{5}\,\delta_{ab}\delta_{cd})\,T_T(u)\right.
\nonumber\\
&\qquad\qquad\quad
\left.+(\delta_{ac}\delta_{bd}-\delta_{ad}\delta_{bc})\,T_A(u) \right)\,.
\label{eq:TTTT}
\end{align}

In the case of four displacement operators, there is a similar decomposition of $SO(3)$ tensor structures 
\begin{align}
\dbra \D_{i}(t_1)\D_{j}(t_2)\D_{k}(t_3)\D_{l}(t_4)\dket
&=\frac{\tilde\gamma^{2}}{t_{12}^{4}\,t_{34}^{4}}
\left(\delta_{ij}\delta_{kl}\,D_S(u)
+(\delta_{ik}\delta_{jl}+\delta_{il}\delta_{jk}
-\tfrac{2}{3}\,\delta_{ij}\delta_{kl})\,D_T(u)\right.
\nonumber\\
&\qquad\quad\qquad +\left.(\delta_{ik}\delta_{jl}-\delta_{il}\delta_{jk})\,D_A(u)\right)\,.
\label{eq:DDDD}
\end{align}

When mixing two tilts with two displacements, there are  two inequivalent
orderings,
\begin{equation}
\dbra \T_a(t_1)\T_b(t_2)\D_i(t_3)\D_j(t_4)\dket
=\frac{\gamma\tilde\gamma \;\delta_{ab}\,\delta_{ij}\;F(u)}{t_{12}^{2}\,t_{34}^{4}}\,,
\label{eq:TTDD}
\end{equation}
as well as
\begin{equation}
\dbra \T_a(t_1)\D_i(t_2)\T_b(t_3)\D_j(t_4)\dket
=\frac{\gamma\tilde\gamma \;\delta_{ab}\,\delta_{ij}\;\tilde F(u)}{t_{13}^{2}\,t_{24}^{4}}\,.
\label{eq:TDTD}
\end{equation}

\subsection{Strong coupling}
\label{app stroncoupling}

At strong coupling all the functions of the cross ratio admit an expansion in powers of
$1/\sqrt\lambda$,
\begin{equation}
T_{S,T,A}=T^{(0)}_{S,T,A}+\tfrac{1}{\sqrt\lambda}\,T^{(1)}_{S,T,A}+\dots\,,
\quad
D_{S,T,A}=D^{(0)}_{S,T,A}+\tfrac{1}{\sqrt\lambda}\,D^{(1)}_{S,T,A}+\dots\,,
\end{equation}
and similarly for $F$ and $\tilde F$.  The leading order (LO) terms come from disconnected contributions, and the $\tfrac1{\sqrt\lambda}$ terms come from the
tree-level contact Witten diagrams of $AdS_2$ worldsheet fluctuations \cite{Giombi:2017cqn},
\begin{align}
T^{(0)}_S&=1+\tfrac{2}{5}T^{(0)}_T\,, &
T^{(0)}_T&=\tfrac12\Big[u^{2}+\tfrac{u^{2}}{(1-u)^{2}}\Big]\,, &
T^{(0)}_A&=\tfrac12\Big[u^{2}-\tfrac{u^{2}}{(1-u)^{2}}\Big]\,,\\
D^{(0)}_S&=1+\tfrac{2}{3}D^{(0)}_T\,, &
D^{(0)}_T&=\tfrac12\Big[u^{4}+\tfrac{u^{4}}{(1-u)^{4}}\Big]\,, &
D^{(0)}_A&=\tfrac12\Big[u^{4}-\tfrac{u^{4}}{(1-u)^{4}}\Big]\,,\\
F^{(0)}&=1\,, & \tilde F^{(0)}&=1\,. &&
\end{align}
The expressions of these functions at next-to-leading order (NLO) become more intricate. Here we quote the functions in the case  of four tilts, as we use them
\begin{align}
T^{(1)}_S&=-\frac{2\left(u^4-4u^3+9u^2-10u+5\right)}{5(u-1)^2}
+\frac{u^{2}\left(2u^4-11u^3+21u^2-20u+10\right)}{5(u-1)^3}\log u
\nonumber\\
&\qquad-\frac{2u^4-5u^3-5u+10}{5u}\log(1-u)\,,
\label{TS1}
\\
T^{(1)}_T&=-\frac{u^{2}\left(2u^2-3u+3\right)}{2(u-1)^2}
+\frac{u^{4}\left(u^2-3u+3\right)}{(u-1)^3}\log u
-u^{3}\log(1-u)\,,
\label{TT1}
\\
T^{(1)}_A&=\frac{u\left(-2u^3+5u^2-3u+2\right)}{2(u-1)^2}
+\frac{u^{3}\left(u^3-4u^2+6u-4\right)}{(u-1)^3}\log u
-\left(u^3-u^2-1\right)\log(1-u)\,,
\label{TA1}
\end{align}
for the case of four displacements, explicit expressions for $D^{(1)}_{S,T,A}$ are also found in \cite{Giombi:2017cqn} and for the mixed correlators,
\begin{equation}
F^{(1)}(u)=-4\left[1-\big(\tfrac12-\tfrac1u\big)\log(1-u)\right]\,,
\quad
\tilde F^{(1)}(u)=-4\left[1+\big(\tfrac12-u\big)\log\left(\tfrac{u}{1-u}\right)\right]\,.
\label{eq:FNLO}
\end{equation}

\subsection{Bose symmetries}

In 1-dimensional CFTs the ordering of operators along the line matters. While $t_1<t_2<t_3<t_4$ is assumed, expressions for the functions in the 4-point correlator are valid in the windows of values $0<u<1$. For configurations with other orderings, the corresponding functions of the cross-ratio are not, in general, given by the same functions, unless the new value for the cross-ratio remains in the windows $0<u<1$ \cite{Ferrero:2023gnu}.

The LO and NLO expressions for the functions of cross-ratio presented in \ref{app stroncoupling} are rather particular in this regard. Witten diagrams in AdS$_2$ enforce the verification of all these Bose symmetries. For example, consider four identical tilt operators
\begin{equation}
    \dbra \T_{1}(t_1)\T_{1}(t_2)\T_{1}(t_3)\T_{1}(t_4)\dket =
    \frac{\gamma^{2} T(u)}{t_{12}^{2}\,t_{34}^{2}}\,,
    \qquad T(u)=T_S(u)+\frac{8}{5}T_T(u)
\end{equation}
The following Bose symmetry conditions are verified for the expressions given in \ref{app stroncoupling},
\begin{align}
    T(1-u) = &\ \left(\tfrac{1-u}{u}\right)^2 T(u)\,,
    \\
     T\left(\tfrac{1}{u}\right) = &\ \left(\tfrac{1}{u}\right)^2 T(u)\,,
     \\
          T\left(\tfrac{u}{u-1}\right) = &\ T(u)\,.
\end{align}
The third one is actually not independent of the first two conditions. For the verification of the last two conditions, one has to replace $\log(1-u)$ by $ \log|1-u|$ and $\log(u) $ by $ \log|u|$ in the NLO expressions. This is needed because the maps $u\to \frac{1}{u}$ and $u\to \frac{u}{u-1}$ take the cross-ratio out of the window $0<u<1$.

For the mixed correlators \eqref{eq:TTDD} and \eqref{eq:TDTD}, the expressions given in \eqref{eq:FNLO} satisfy the Bose symmetries associated with the transposition of two identical operators
\begin{align}
F\left(\tfrac{u}{u-1}\right) = & \ F(u)\,,
\label{BoseF}
\\
\tilde F(1-u)= & \ \tilde F(u)\,.
\label{BoseFt}
\end{align}

Finally, the transposition $(2,3)$ in the mixed correlators exchanges a tilt with a displacement, which maps the ordering of \eqref{eq:TTDD} into that of \eqref{eq:TDTD}.
What one obtains in this case is a relation between the two different 
functions
\begin{equation}
F\left(\tfrac{1}{u}\right) = \tilde F(u)\,.
\label{BoseFtF}
\end{equation}

\section{Fluctuations on the world-sheet}
\label{app:fluc}
Here, we review some details of the spectrum of bosonic quadratic fluctuations on the open string on $AdS_5\times S^5$ dual to a Wilson line with both geometric and internal cusp angles \cite{Drukker:2007qr, Drukker:2011za}.

The classical open string extends on a subspace $AdS_3 \times S^1 \subset AdS_5 \times S^5$, whose metric is given by
\begin{equation}
    ds^2=R^2\left(\cosh^2\rho\, d\uptau^2+d\rho^2+\sinh^2\rho d\varphi^2+d\vartheta^2\right)\,,
\end{equation}
when working with euclidean signature. The Nambu-Goto equations of motion are solved using a static gauge ansatz relating $(\uptau,\vartheta)$ with world-sheet coordinates $(\tau,\sigma)$ as follows,

\begin{alignat}{2} 
 \uptau &= b \sqrt{\frac{1-k^2}{1+b^2}}\,\tau\,, & \quad \vartheta &= \sqrt{\frac{1-(2+b^2)k^2}{1+b^2}}\, \sigma \\ 
 \cosh \rho(\sigma) &= \frac{\sqrt{1+b^2}}{b}\frac{1}{{\rm cn}(\sigma,k^2)}\,, & \qquad \varphi'(\sigma) &= \frac{1}{b}\sqrt{\frac{b^2+k^2}{1+b^2}} \frac{1}{\sinh^2\rho(\sigma)}\,,
\end{alignat}
This solution is written in terms of 2 constants of motion $(b,k)$ that eventually parametrize the cusp angles of the dual Wilson line. The induced metric on the world-sheets
\begin{equation}
    ds_{\text{ind}}^2 =  \frac{R^2(1 - k^2)}{\text{cn}^2(\sigma,k^2)} \left(d\tau^2 + d\sigma^2 \right)\,,
    \label{eq:equation_label}
\end{equation}
with ranges
\begin{equation}
    -\mathbb{K}(k^2)<\sigma<\mathbb{K}(k^2)\,, \qquad -\infty<\tau<\infty\,.
\end{equation}

In the classical string solution, the values of the cusp angles $\phi$ and $\theta$ are captured by the total spanning of the functions $\varphi(\sigma)$ and $\vartheta(\sigma)$. In terms of the
parameters $b$ and $k$
\begin{eqnarray}
    \phi &\!=\!& \pi-2\frac{b(1-k^2)}{\sqrt{(1+b^2)(b^2+k^2)}}\left[\Pi\left(\tfrac{b^2+k^2}{1+b^2}|k^2 \right)-\mathbb{K}(k^2) \right]
    \label{cuspgeometrico}\,,
    \\
    \theta&\!=\!&
   2 \sqrt{\frac{1-k^2(2+b^2)}{1+b^2}}
   \mathbb{K}(k^2)\,.
    \label{cuspinterno}
\end{eqnarray}

The expansion around this classical solution up to quadratic order is of the form
\begin{equation}
    S_{\rm fluc}^{(\phi,\theta)} = \frac{\sqrt\lambda}{4\pi}\int \! d\tau\! \int^{\mathbb{K}}_{-\mathbb{K}}\! d\sigma\,\sqrt{g} \left(\partial^\mu\varphi_\alpha \partial_\mu\varphi_\alpha + M_{\alpha\beta}(\sigma) \varphi_\alpha \varphi_\beta + A(\sigma)(\varphi_3\partial_{\sigma}\varphi_8-\varphi_8\partial_{\sigma}\varphi_3)\right), 
\end{equation}
The complete set of coupling parameters can be found in \cite{Drukker:2011za}. The simplest ones are,

\begin{eqnarray}
    M^2_{\alpha\alpha} &\!=\!& 2+ \frac{k^2(2+b^2)-1}{(1+b^2)(1-k^2)}{\rm cn}^2(\sigma,k^2), \qquad \alpha =  1, 2\,,
    \\
    M^2_{\alpha\alpha} &\!=\!& \frac{k^2(2+b^2)-1}{(1+b^2)(1-k^2)}{\rm cn}^2(\sigma,k^2), \qquad \alpha =  4, 5, 6, 7\,.
\end{eqnarray}

\section{Green's Functions}
\label{app green}
In this section we construct the bulk-to-bulk propagator of the fluctuations with the equation of motion \eqref{simpleeom} and use it to re-obtain the expression of boundary-to-boundary propagators \eqref{fourier1} and \eqref{fourier2} in configuration space.

The kind of fluctuations we consider obey the equation of motion
\begin{equation}
    \nabla_\mu\nabla^\mu \varphi_\alpha + \frac{1}{\sqrt{g}}  M^2
    \varphi_\alpha = 0\,,
    \label{simpleeom2}
\end{equation}
where $M$ is some constant related to the internal cusp angle according to $\theta^2=4 \mathbb{K}(k^2)^2M^2$. The bulk-to-bulk propagator $G$ is the solution of the corresponding equation with a $\delta$-function source. In the case of interest to us, we have to impose as 
 boundary conditions that it vanishes at $\sigma\to\pm \mathbb K$. In the conformal gauge \eqref{induced}, $\sqrt{g}g^{a b}=\delta^{a b}$ and this equation can be rewritten as
\begin{equation}
\Big(\partial_\tau^2+\partial_\sigma^2+\frac{\theta^2}{4\mathbb{K}^2}\Big)\,
G(\tau,\sigma;\tau',\sigma')=\delta(\tau-\tau')\,\delta(\sigma-\sigma')\,.
\label{eq:Geq}
\end{equation}
Green's identity enables one to express the solution of \eqref{simpleeom2} with boundary conditions $\varphi_\alpha(\tau,\sigma_I)=A_\alpha^I(\tau)$.
Denoting by $\partial_n$ the outward normal derivative, we have $\partial_n=\partial_\sigma$ at $\sigma=\mathbb{K}$ and $\partial_n=-\partial_\sigma$ at $\sigma=-\mathbb{K}$, and the solution reads
\begin{equation}
\varphi_\alpha(\tau,\sigma)= 
\sum_{I={\rm L,R}}\int d\tau'\,A^I_\alpha(\tau')\,\partial_{n'}G(\tau,\sigma;\tau',\sigma_I)\,,
\qquad \sigma_{\rm L}=-\mathbb{K},\ \ \sigma_{\rm R}=\mathbb{K}\,.
\end{equation}
From the near-boundary expansions of the fields, $B^I_\alpha=-\partial_n\varphi_\alpha|_{\sigma_I}$, so that $\delta B^I(\tau)/\delta A^J(\tau')=-\partial_n\partial_{n'}G(\tau,\sigma_I;\tau',\sigma_J)$. Together with $\delta S_{\rm fluc}^{(\phi,\theta)}=-\frac{\sqrt\lambda}{2\pi}\int d\tau\,\delta A^I B^I$ this gives
\begin{equation}
    -\left.\frac{\delta^2 S_{fluc}^{(\phi,\theta)}}{\delta A_\alpha^I (\tau)\delta A_\alpha^J (\tau')}\right|_{A=0}=  -\frac{\sqrt{\lambda}}{2 \pi} \partial_n \partial_{n'} G(\tau,\sigma_I;\tau',\sigma_J),\ \ \ \ \ 
    I,J = {\rm L}, {\rm R}\,.
    \label{ddG}
\end{equation}

For the case $\theta=0$ one can find a closed form solution of the propagator. The starting point is the Green's function in the whole plane,
\begin{equation}
    G(\tau,\sigma;\tau',\sigma')=\frac{1}{4\pi}\log{\[(\tau-\tau')^2+(\sigma-\sigma')^2\]}\,.
\end{equation}
Through the standard method of images, one can find the Green's function for the strip $-\mathbb{K}<\sigma<\mathbb{K}$, that vanishes as $\sigma,\sigma'\to \pm {\mathbb{K}}$
\begin{equation}
G(\tau,\sigma;\tau',\sigma')=\frac{1}{4\pi}\,
\log\left[\frac{\cosh\!\big(\frac{\pi}{2\mathbb{K}}(\tau-\tau')\big)
          -\cos\!\big(\frac{\pi}{2\mathbb{K}}(\sigma-\sigma')\big)}
         {\cosh\!\big(\frac{\pi}{2\mathbb{K}}(\tau-\tau')\big)
          +\cos\!\big(\frac{\pi}{2\mathbb{K}}(\sigma+\sigma')\big)}\right],
\quad -\mathbb{K}<\sigma,\sigma'<\mathbb{K}.
\label{eq:G0}
\end{equation}
From the bulk-to-bulk propagator one can also verify that
\begin{equation}
 -\frac{\sqrt{\lambda}}{2\pi}\,\partial_{n'}\partial_{n}G\,\bigg|_{\substack{\sigma=\mathbb{K}\\ \sigma'=\mathbb{K}}}=-\frac{\sqrt{\lambda}}{2\pi}\,\partial_{n'}\partial_{n}G\,\bigg|_{\substack{\sigma=-\mathbb{K}\\ \sigma'=-\mathbb{K}}}=\frac{\sqrt\lambda}{32\mathbb{K}(k^2)^2}\frac{1}{\sinh^2\left(\frac{\pi(\tau'-\tau)}{4\mathbb{K}(k^2)}\right)}
\end{equation}
in agreement with \eqref{sinh} and that
\begin{equation}
 -\frac{\sqrt{\lambda}}{2\pi}\,\partial_{n'}\partial_{n}G\,\bigg|_{\substack{\sigma=-\mathbb{K}\\ \sigma'=\mathbb{K}}}=-\frac{\sqrt{\lambda}}{2\pi}\,\partial_{n'}\partial_{n}G\,\bigg|_{\substack{\sigma=\mathbb{K}\\ \sigma'=-\mathbb{K}}}=\frac{\sqrt\lambda }{32\mathbb{K}(k^2)^2}\frac{1}{\cosh^2\left(\frac{\pi(\tau'-\tau)}{4\mathbb{K}(k^2)}\right)}
\end{equation}
in agreement with \eqref{cosh}. 

For arbitrary values of $\theta$ we can solve \eqref{eq:Geq} by expanding the Green's function $\sigma$ dependence in terms of Dirichlet eigenfunctions on the interval which satisfy
\begin{equation}
\partial_\sigma^2 u_m(\sigma)=-\left(\frac{m \pi}{2 \mathbb{K}}\right)^2 u_m(\sigma), \quad\Rightarrow\quad
u_m(\sigma)=\frac{1}{\sqrt{\mathbb{K}}}\sin \left[\frac{m \pi (\sigma+\mathbb{K})}{2 \mathbb{K}} \right],\quad m=1,2,\dots
\end{equation}
These modes vanish at $\sigma=\pm{\mathbb{K}}$ and have been normalized such that

\begin{equation}
   \int_{-\mathbb{K}}^\mathbb{K} u_m(\sigma) u_{m'}(\sigma)=\delta_{m m'}\,, \qquad \sum_{m=1}^\infty u_m(\sigma)u_m(\sigma')=\delta(\sigma-\sigma')\,.
\end{equation}

For the $\tau$-dependence, we use $g_m(\tau)$ which obeys
\begin{equation}
 (\partial_\tau^2-\omega_m^2)g_m=\delta(\tau-\tau')\quad{\rm for}\ \    
 \omega_m^2=\frac{m^2 \pi^2-\theta^2}{4 \mathbb{K}^2}\quad
 \Rightarrow\quad
 g_m(\tau)=-\frac{e^{-\omega_m|\tau|}}{2\omega_m}\,.
\end{equation}

Then, the solution to the Green's function equation \eqref{eq:Geq} is
\begin{align}
    G(\tau,\sigma;\tau',\sigma')= & \ \sum_m u_m(\sigma) u_m(\sigma')g_m(\tau-\tau') 
    \nonumber
    \\
    = &\ -\sum_{m=1}^\infty \frac{\sin \left[\frac{m \pi (\sigma+\mathbb{K})}{2 \mathbb{K}} \right]\sin \left[\frac{m \pi (\sigma'+\mathbb{K})}{2 \mathbb{K}} \right]}{\sqrt{m^2 \pi^2-\theta^2}}  e^{-\sqrt{m^2 \pi^2-\theta^2}\frac{|\tau-\tau'|}{2\mathbb{K}}}
\end{align}
Using \eqref{ddG} and that $\left.\partial_n u_m\right|_{\sigma=-\mathbb{K}}=-\frac{m \pi}{2 \mathbb{K}^{3/2}}$ and $\left.\partial_n u_m\right|_{\sigma=\mathbb{K}}=(-1)^{m}\frac{m \pi}{2 \mathbb{K}^{3/2}}$ we obtain: 
\begin{align}
    -\left.\frac{\delta^2 S_{fluc}^{(\phi,\theta)}}{\delta A_\alpha^R (\tau)\delta A_\alpha^R (\tau')}\right|_{A=0} = &\ \frac{\sqrt{\lambda}}{8 \pi \mathbb{K}^2}  \sum_{m=1}^\infty \frac{\pi^2 m^2} {\sqrt{m^2 \pi^2-\theta^2}}  e^{-\sqrt{m^2 \pi^2-\theta^2}\frac{|\tau-\tau'|}{2\mathbb{K}}},
\\
    -\left.\frac{\delta^2 S_{fluc}^{(\phi,\theta)}}{\delta A_\alpha^L (\tau)\delta A_\alpha^R (\tau')}\right|_{A=0} = &\ \frac{\sqrt{\lambda}}{8 \pi \mathbb{K}^2} \sum_{m=1}^\infty \frac{(-1)^{m+1}\pi^2 m^2} {\sqrt{m^2 \pi^2-\theta^2}}  e^{-\sqrt{m^2 \pi^2-\theta^2}\frac{|\tau-\tau'|}{2\mathbb{K}}},
\end{align}
which exactly reproduces the equations \eqref{fourier1solt} and \eqref{fourier2solt}.

\end{document}